\documentclass[11pt,a4paper]{article}
\usepackage[utf8]{inputenc}
\usepackage{amsmath}
\usepackage{amsfonts}
\usepackage{amssymb}
\usepackage{amsthm} 
\usepackage{graphicx}
\usepackage[table, usenames, dvipsnames]{xcolor}
\usepackage{placeins} 
\usepackage[version=3]{mhchem} 
\usepackage{url}

\usepackage[breaklinks]{hyperref}
\usepackage{doi}
\usepackage{settobox} 
\usepackage{algorithm}
\usepackage{algpseudocodex} 
\usepackage{lscape} 
\usepackage{multirow} 
\usepackage{longtable} 
\usepackage{comment} 
\usepackage{colortbl} 
\usepackage{amsopn}
\usepackage{bm}
\usepackage{cleveref} 
\usepackage{subfig} 

\newbox{\subbox}
\newlength{\xheight}
\newlength{\subheight}

\usepackage[margin=3cm]{geometry}
\usepackage[sort&compress,comma,square,numbers]{natbib} 

\theoremstyle{plain}
\newtheorem{thm}{Theorem}[section]

\theoremstyle{definition}
\newtheorem{definition}[thm]{Definition}
\newtheorem{remark}[thm]{Remark}

\usepackage{multirow}

\newcommand{\res}{{\rm res}}
\newcommand{\degree}{{\rm degree}}
\renewcommand{\algorithmicrequire}{\textbf{Input:}}
\renewcommand{\algorithmicensure}{\textbf{Output:}}
\newcommand\Inputo{\Statex \algorithmicrequire\ }
\newcommand\Outputo{\Statex \algorithmicensure\ }
\newcommand{\R}{\mathbb{R}}
\newcommand{\lv}{\rm{lv}}
\newcommand{\bp}{\rm{BP}}

\newcommand{\footremember}[2]{%
	\footnote{#2}
	\newcounter{#1}
	\setcounter{#1}{\value{footnote}}%
}
\newcommand{\footrecall}[1]{%
	\footnotemark[\value{#1}]%
} 

\author{AmirHosein Sadeghimanesh\footremember{this}{Research Centre for Computational Sciences and Mathematical Modelling, Coventry University, Coventry, United Kingdom.} \footremember{Contact-AS}{amirhossein.sadeghimanesh@coventry.ac.uk.}\footremember{corr}{Corresponding author.} and Matthew England\footrecall{this} 
}

\title{Resultant Tools for Parametric Polynomial Systems with Application to Mathematical Biology}

\begin{document}
	
	\maketitle

\begin{abstract}
	To decompose the parameter space of a parametric system of polynomial equations with respect to the number of real solutions that the system attains, first the discriminant variety of the system should be computed. This is an elimination type problem because the discriminant variety is the union of the projections into the parameter space of the intersection of the variety of the original system with one or several new hypersurfaces. Despite the popularity of Gr\"obner Bases (GBs) for this task, in this work we focus on a lesser used alternative tool, namely resultants.
	
	We develop new approaches to build the discriminant variety using different resultant methods such as Dixon resultant and iterated univariate resultants. We discuss how the minimal discriminant variety, the discriminant varieties computed by our methods, and the known border polynomials are related. We demonstrate with numerous examples from population dynamics and chemical reaction network theory that our new approach successfully handles larger size examples than a GB can.
	
\end{abstract}

\subsection{Problem statement}
\label{sec:Problem_Statement}

Let $R=\mathbb{Q}[k_1,\dots,k_r][x_1,\dots,x_n]$ be the ring of polynomials in $n$ variables $x=(x_1,\dots,x_n)$ with coefficients coming from the ring of polynomials in $r$ parameters $k=(k_1,\dots,k_r)$ with rational coefficients. A system of parametric polynomial equations is defined by a finite set of polynomials $F=\lbrace f_1,\dots,f_m\rbrace\subset R$. For each specification of the parameters $k^\star\in\mathbb{R}^r$, the solution set to $f_1=\dots=f_m=0$ will be a subset of $\mathbb{R}^n$, i.e. the variety $V(F|_{k=k*})$.

In many applications the topic of study can be modelled by such a parametric system of polynomial equations.  It is often the case in such applications that the system of interest has finitely many solutions (for generic choices of the parameters). In such cases one common desire is to know the possible number of solutions and the parameter regions where each of these possible numbers are attained.  This is the case for example in chemical reaction network theory (see e.g. \cite{BDEEGGHKRSW20}) and for the study of population dynamics (see e.g. \cite{Knipl-Rost-2014}).

Thus the problem we study is how to decompose the parameter space for a system of parametric polynomial equation into connected subsets, where the number of solutions to a polynomial system is invariant.  

\subsection{Decompositions}

One potential tool is Cylindrical Algebraic Decomposition (CAD). Invented by Collins in the 1970s \cite{Collins-1975}, CAD produces a decomposition of an $N$-dimensional real space $\mathbb{R}^N$ into connected components (cells) which are semi-algebraic (may be described by a sequence of polynomial constraints). The cells are cylindrical with respect to a given variable ordering: meaning the projections of any two cells onto a lower coordinate space in the same ordering are either equal or disjoint.  I.e. the cells stack up in cylinders.  

Collins' original CAD algorithm produced cells on which a set of input polynomials were all sign-invariant (i.e. positive, zero, or negative throughout a given cell).  One can then check a single sample point of a cell and infer many properties throughout the cell, such as the truth of any formulae built with the polynomials.  The original motivation of Collins was to allow for quantifier elimination over the reals \cite{Collins-1975}.  The common framework of most CAD algorithms involves two stages: first a projection stage to progressively identify polynomials of fewer variables, and then a lifting stage which uses these polynomials to build the decomposition.


A decomposition of the parameter space sign-invariant for the polynomials in our input equations would meet our requirements, however, it would likely involve far more cells than needed.  Since its inception, CAD has been developed intensively, with one path of improvements on invariance properties weaker than sign invariance but still sufficient for the problem at hand \cite{Bradford-Davenport-England-McCallum-Wilson-Projection-Doubly-exponential,McCallum-1999}.  However, CAD has complexity doubly exponential in $N$ \cite{BD07}.  In the context of our problem, $N$ is the total number of indeterminates (i.e. both variables $x$ and parameters $k$), thus doubly exponential in $r+n$.  So although CAD is suited to the problem in theory, it is not practical as a tool on its own.  

Recall that we want a decomposition of only the parameter space.  Thus we may simplify CAD to perform the full projection and terminate lifting once the parameter space alone is decomposed.  However, the doubly exponential complexity occurs already in the projection stage:  this approach will still provide something more fine-grained than our requirement of invariance for the number of solutions to the system.  

\subsection{Contributions and plan of the paper}
\label{sec:Plan_of_the_Paper}

Hope lies in the combination of CAD with other algebraic approaches.  For example, when CAD was used for chemical reaction network analysis in \cite{BDEEGGHKRSW20} it was combined with virtual term substitution and lazy real triangularisation.  
The present state-of-the-art for our problem is generally considered to be a combination of CAD with another tool: the discriminant variety \cite{Lazard-Rouillier-2007}.  This is described by polynomials in the parameters and provides the boundaries between the invariant regions we seek. We may first compute this and then perform a sign-invariant CAD of only the parameter space with respect to it; an approach described in \Cref{sec:Solution_via_CAD_and_Discriminant_Variety}.  
However, this has proven infeasible for recent studies of population models \cite{Rost-Sadeghimanesh-2021-1}.  We introduce those models next in \Cref{sec:Population_Models_with_the_Allee_Effect} and then later in \Cref{sec:Recent_Prior_work_on_Population_Model_Application} we summarise the recent attempts which had to resort to symbolic-numeric methods. 

We then describe our new contributions in \Cref{sec:New_Approach_using_Resultants}, which allow for a purely symbolic solution to this problem. The new approaches replace the traditional Gr\"obner basis method to obtain the discriminant variety with resultant techniques.  Resultants are an ingredient of the CAD projection stage, but our use of them is less extensive than CAD. The symbolic solution to the population model problem is then described in \Cref{sec:Application_of_New_Approach_to_Population_Model}. Section~\ref{sec:Other_groups_attempts} discusses works of other groups on this population model.

Section~\ref{sec:Benchmarking} contains further benchmarking on models from both population dynamics and Chemical Reaction Network (CRN) theory to compare performance of our new algorithms with the currently commonly used ones. It also shows that our algorithms work in a generic setting and are not designed for a narrow class of biological models.

\subsection{Notation}
\label{sec:notations}

A set is represented by braces, e.g. $\lbrace a,b,c\rbrace$, and by an ordered set we mean a totally ordered set. An ordered set is represented by round parenthesis, for example a set $\lbrace a,b,c\rbrace$ with the relation $a<b<c$ is shown as $(a,b,c)$. An empty ordered set is denoted by $()$. If $N_1>N_2$ are two positive integers, then the projection map $\pi\colon\R^{N_1}\rightarrow\R^{N_2}$ that sends any points $(a_1,\dots,a_{N_1})$ to $(a_1,\dots,a_{N_2})$, i.e. discarding the last $N_1-N_2$ coordinates, is denoted by $\pi_{N_2}^{N_1}$. The image of a set $A\subseteq\R^{N_1}$ under this map is denoted by $\pi_{N_2}^{N_1}(A)$.

\subsection{Computational Environment}
\label{sec:Computational_environment}

All reported computations in this paper were performed on a Windows 11 machine with Intel(R) Core(TM) Ultra 7 155H, 3800 Mhz, 16 Core(s), 22 Logical Processor(s), and 32.0 GB (RAM). The versions of the software and programming language used were Maple 2025.2 and Python 3.13.1.

\section{Population Models -- Allee Effect}
\label{sec:Population_Models_with_the_Allee_Effect}

A well-known population model is \emph{logistic growth}, where due to a limitation of resources the population can not exceed a certain level. At the beginning, when the size of population is small, because of an abundance of resources the growth of the population is high; but as time passes and the population increases, the amount of available resources per individual decreases and the speed of growth reduces until eventually the population reaches a steady state, called the carrying capacity of the system \cite{Logistic-population-growth-2020}.  

The \emph{Allee effect} is a less well-known phenomenon in biology where the population is not only competing for resources, as in logistic growth models, but also has cooperative behaviour which increases the chance of survival. The Allee effect was first described by an American ecologist, Warder Clyde Allee in the 1930s when he was studying the behaviour of a goldfish population \cite{Allee-1932}.

A \emph{strong Allee effect} happens when the population needs to be above a certain amount, called the Allee threshold, to benefit from the cooperative behaviour and be safe from extinction. An Allee effect can be caused by different reasons, e.g. at a low population density the species has difficulty finding mates for reproduction and fertilisation \cite{Strong-Allee-example-Fig-trees}. A strong Allee effect behaviour has been observed in various species such as starfishes \cite{Stong-Allee-effect-example-starfish-1,Stong-Allee-effect-example-starfish-2}, bacteria \cite{Strong-Allee-effect-example-bacteria} and honey bees \cite{Zeaiter-Myerscough-2020}.

\begin{figure}[ht]
	\begin{center}
		\includegraphics[width=5cm]{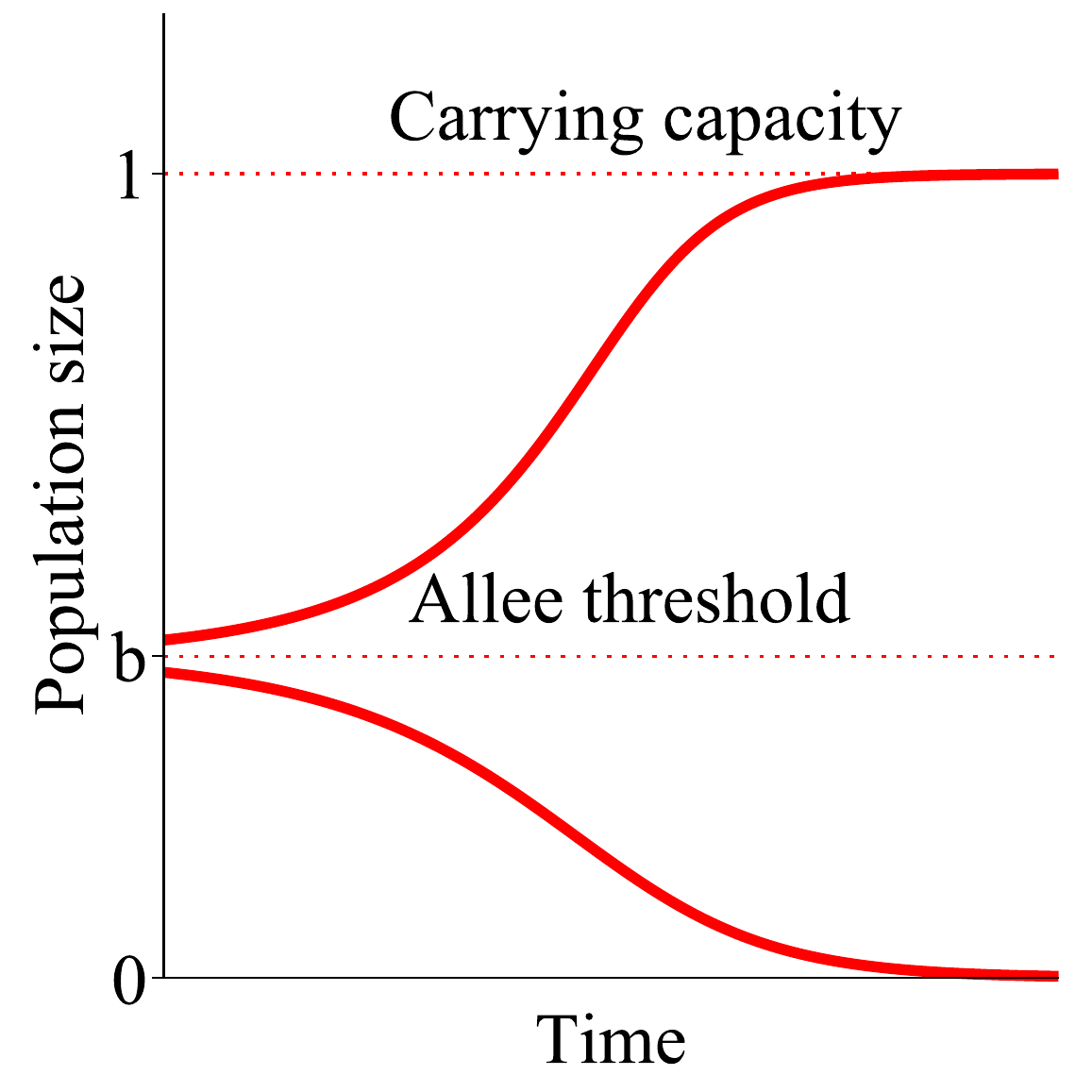}
	\end{center}
	\caption{Dynamical behaviour of a single population with the strong Allee effect. If the initial size of the population is below the Allee threshold, the population will eventually become extinct. Otherwise it will survive and increase until it reaches the carrying capacity.}\label{Fig:Allee_effect_normalised}
\end{figure}

A simple population model with the strong Allee effect is:
\[
\frac{dx(t)}{dt}=x(t)\big(1-x(t)\big)\big(x(t)-b\big),
\]
where $x(t)$ is the population size at time $t$.  In this example the carrying capacity is 1 and the Allee threshold is $b$ (where $0<b<1$).  From here on we drop the emphasis on $t$ and write $x$ and $\dot{x}$ instead of $x(t)$ and $dx(t)/dt$.  The dynamical behaviour of a single population with the strong Allee effect is shown in \Cref{Fig:Allee_effect_normalised}. We can easily identify the three steady states by setting the derivative to zero.  Two of them are stable, extinction and the carrying capacity, while a third, the Allee threshold, is unstable. 

The situation becomes more interesting when several populations of the same species with strong Allee effect are connected. The study of dynamical behaviour of connected populations with Allee effects is an ongoing recent research topic \cite{Gyllenberg-Hemminki-Tammaru-1999,Knipl-Rost-2014,Knipl-Rost-2016,Rost-Sadeghimanesh-2021-1,Rost-Sadeghimanesh-2021-2,Vortkamp-Schreiber-Hastings-Hilker-2020}. Consider $n$ populations for $n\in\mathbb{N}$ and denote the size of the $i$-th population with $x_i$. The simplest scenario is to connect all populations to each other with a complete digraph and assume the same dispersal rate for each path. Let $a$ be the dispersal rate and assume that all populations have the same Allee threshold, $b$. The ordinary differential equation system governing the dynamical behaviour of this model is then as follows \cite[Equation~(3)]{Rost-Sadeghimanesh-2021-1}:
\begin{equation}\label{Eq:n-patch-ODE}
	\dot{x}_i=x_i(1-x_i)(x_i-b)-(n-1)ax_i+\sum_{\substack{j=1\\j\neq i}}^nax_j,\quad i=1,\dots,n.
\end{equation}
To study the steady states of this model, one has to obtain the non-negative real solutions to the parametric polynomial system of equations obtained by setting all $\dot{x_i}$ equal to zero in System \eqref{Eq:n-patch-ODE}. Here $a$ and $b$ are the parameters, which may be chosen from $\mathbb{R}_{\geq 0}$, and the $x_i$'s are variables. Of most interest is understanding the different parameter regions which give rise to different numbers of steady states, i.e. a problem as defined in \Cref{sec:Problem_Statement}.

\section{CAD and Discriminant Variety}
\label{sec:Solution_via_CAD_and_Discriminant_Variety}

In this section we describe an approach to tackle the problem introduced in \Cref{sec:Problem_Statement}.  It is described originally in \cite{Lazard-Rouillier-2007,Moroz-PhD-thesis} and is implemented as \texttt{RootFinding:-Parametric} \cite{RootFinding-package} in the computer algebra system Maple.

\subsection{Method and example}\label{sec:Method_and_example}

The method has two steps: the first consists of finding the possible candidates for the boundaries between different regions where the number of solutions may vary; and the second consists of decomposing the parameter space with respect to these possible boundaries in an algorithmic way. Note that in applications in biology one is only really interested in open regions. The reason is that in experiments one can not guarantee to fix the parameters to an exact value due to natural perturbations or inevitable small errors in measurement tools. Therefore we will ask this algorithm to only return the open regions in the decomposition. 

We illustrate on a simple worked example. Let $R=\mathbb{Q}[b,c][x]$ and $F=\lbrace f\rbrace$ where $f=x^2+bx+c$. The parameter space is $\mathbb{R}^2$. For every choice of $(b,c)\in\mathbb{R}^2$ the set $V(F)$ has finitely many real solutions. For the number of points in $V(F)$ to change, one must vary $(b,c)$ such that they cross a value where a solution gets multiplicity higher than one.  This is because to reduce the number of real solutions, two real solutions must collide and leave the real line, and to increase the number of real solutions, two non-real complex solutions must collide and enter the real line. In such situations an additional equation, $df/dx=0$ should hold in addition to $f=0$.  So the first step in this algorithm is to find values for $(b,c)$ such that there exists a choice of $x$ that satisfies $f=df/dx=0$. 

\begin{figure}[ht]
	\centering
	\subfloat[]{\label{Fig:Degree_2_equation_discriminant_real}\includegraphics[width=4.3cm]{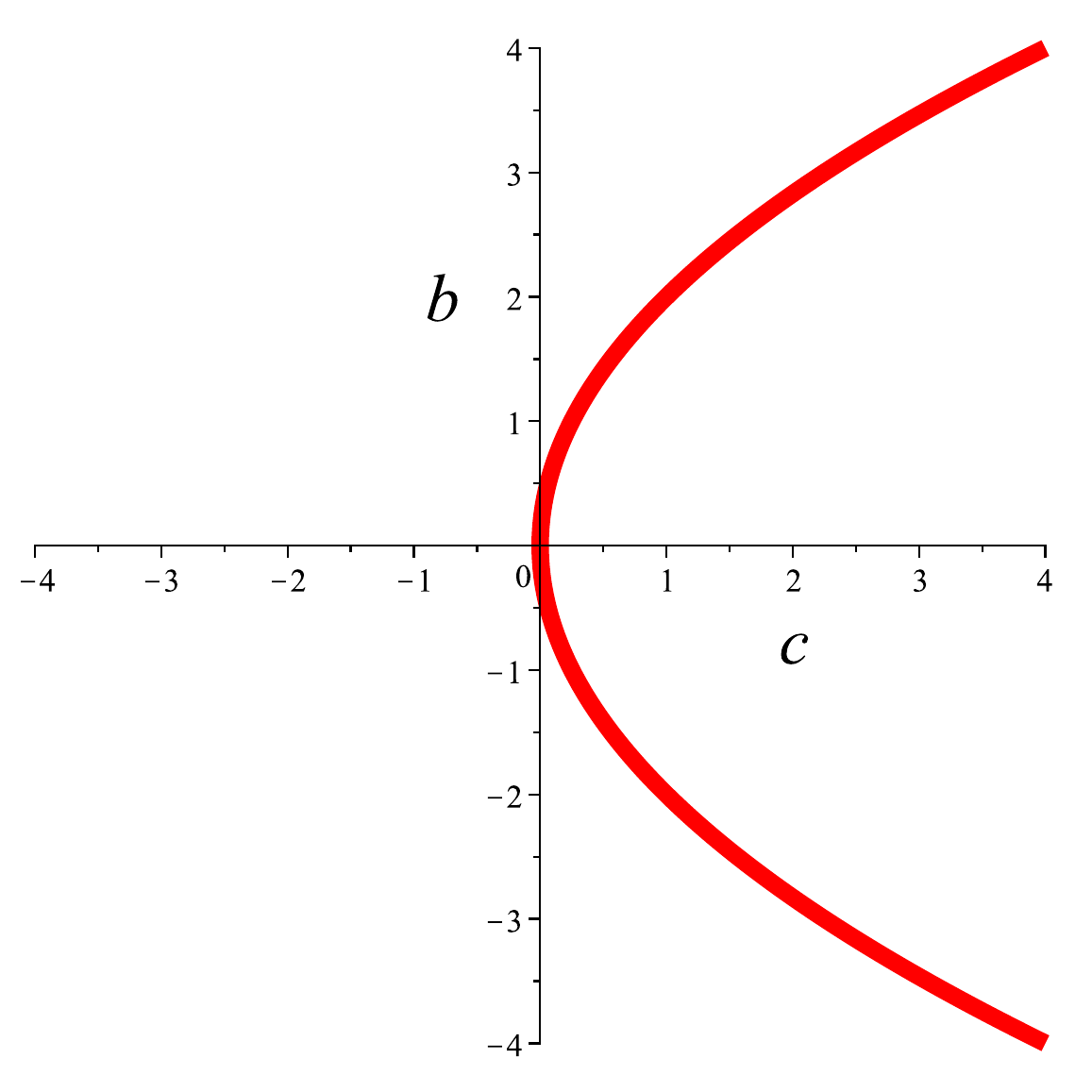}}
	\subfloat[]{\label{Fig:Degree_2_equation_open_CAD}\includegraphics[width=4.3cm]{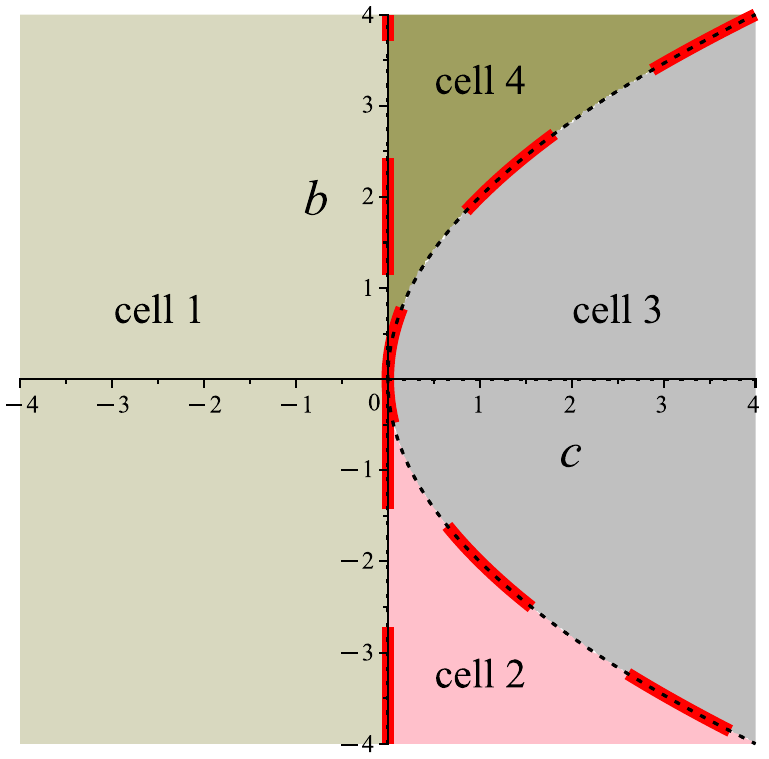}}
	\subfloat[]{\label{Fig:Degree_2_equation_open_CAD_real}\includegraphics[width=4.3cm]{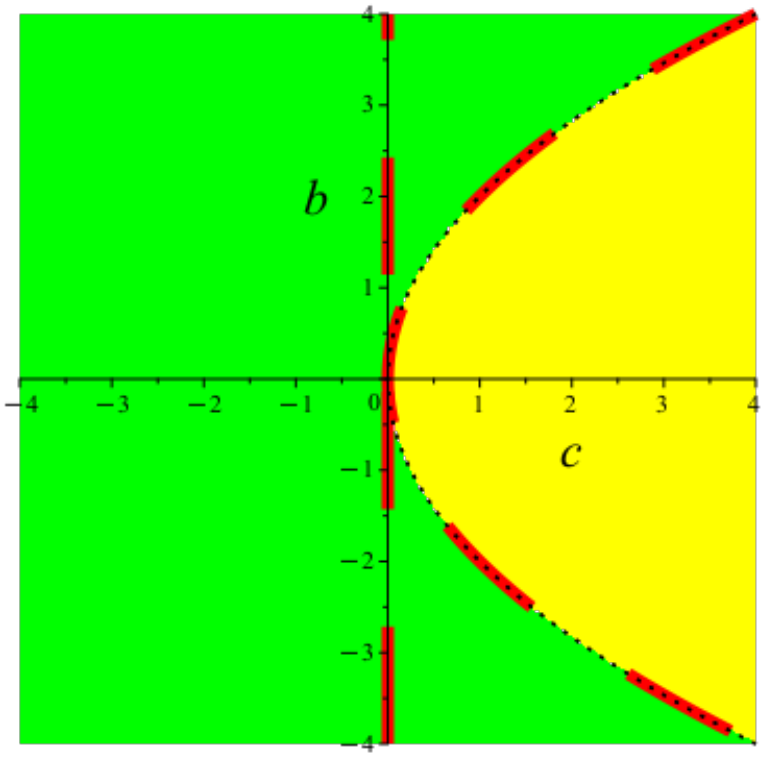}}
	\caption{(a) The discriminant variety of the parametric system $x^2+bx+c=0$ when one studies the number of real solutions. (b) The open CAD of the parameter plane with respect to the discriminant variety in (a).   (c) The system has two real solutions in the green coloured region and no real solution in the yellow coloured region.}
	\label{Fig:Degree_2_equation}
\end{figure}

This is a type of quantifier elimination problem and can be solved using elimination theory, e.g. via Gr\"obner Basis (GB) computation\footnote{For a brief introduction to GB see \cite{Sturmfels2005}, or for an English reproduction of the original thesis which defined GB see \cite{Buchberger2006}.}. We compute a GB for the ideal generated by $x^2+bx+c$ and $2x+b$ with a lexicographic monomial order where we consider $x$ greater than $b$ and $c$ in the ring $\mathbb{Q}[b,c,x]$. We then remove the polynomials involving $x$ from the output. The variety of the result is the Zariski closure of the set of parameters that we are looking for. In this example we end up with a single polynomial $4c-b^2$. This is the discriminant and more generally the output of the process is called the \emph{discriminant variety}. \Cref{Fig:Degree_2_equation_discriminant_real} shows $V(4c-b^2)$. 

The second step of the method is to use an open CAD (i.e. a CAD which returns only the cells of full dimension \cite{WBDE14}) to decompose $\mathbb{R}^2$ with respect to this curve. This will produce four open sets for our example, as shown in \Cref{Fig:Degree_2_equation_open_CAD}. The number of real points in $V(F)$ is invariant over each of these open cells. Therefore, it is now enough to pick one sample point from each, and then repeatedly solve the system and count the number of real solutions after substituting each of these values for the parameters of the system. 

In this example, Cells 1, 2 and 4 have two real solutions and in Cell 3 the system has no real solution.   \Cref{Fig:Degree_2_equation_open_CAD_real} shows the result of this analysis.  Here the regions of the same colour indicate the same number of solutions of the system.  We acknowledge that the number of solutions on the boundaries of the decomposition (the dashed lines) are not determined by this process, but if desired they could be uncovered by computing the full CAD and testing the additional lower dimensional cells.

\subsection{Restricting to positive solutions}\label{sec:Restricting_to_positive_solutions}

In many applications it is common to care only about the positive solutions to the system (it is not possible to have a population of negative size for example).  This means that in addition to our parametric polynomial equations we have inequalities also.  Suppose we include $x>0$ along with $f=0$ from our simple example. In this case the discriminant variety has an extra component. Note that the sign of the real solutions may change if by varying the parameters we cross a choice of $(b,c)$ such that a solution to the system becomes zero. So again we have a new quantifier elimination problem. We want to check if there are choices of $(b,c)$ such that there exists $x$ satisfying $f=x=0$. Once again, we can solve this using a GB computation to obtain the polynomial $c$. 

\begin{figure}[ht]
	\centering
	\subfloat[]{\label{Fig:Degree_2_equation_discriminant_positive}\includegraphics[width=4.5cm]{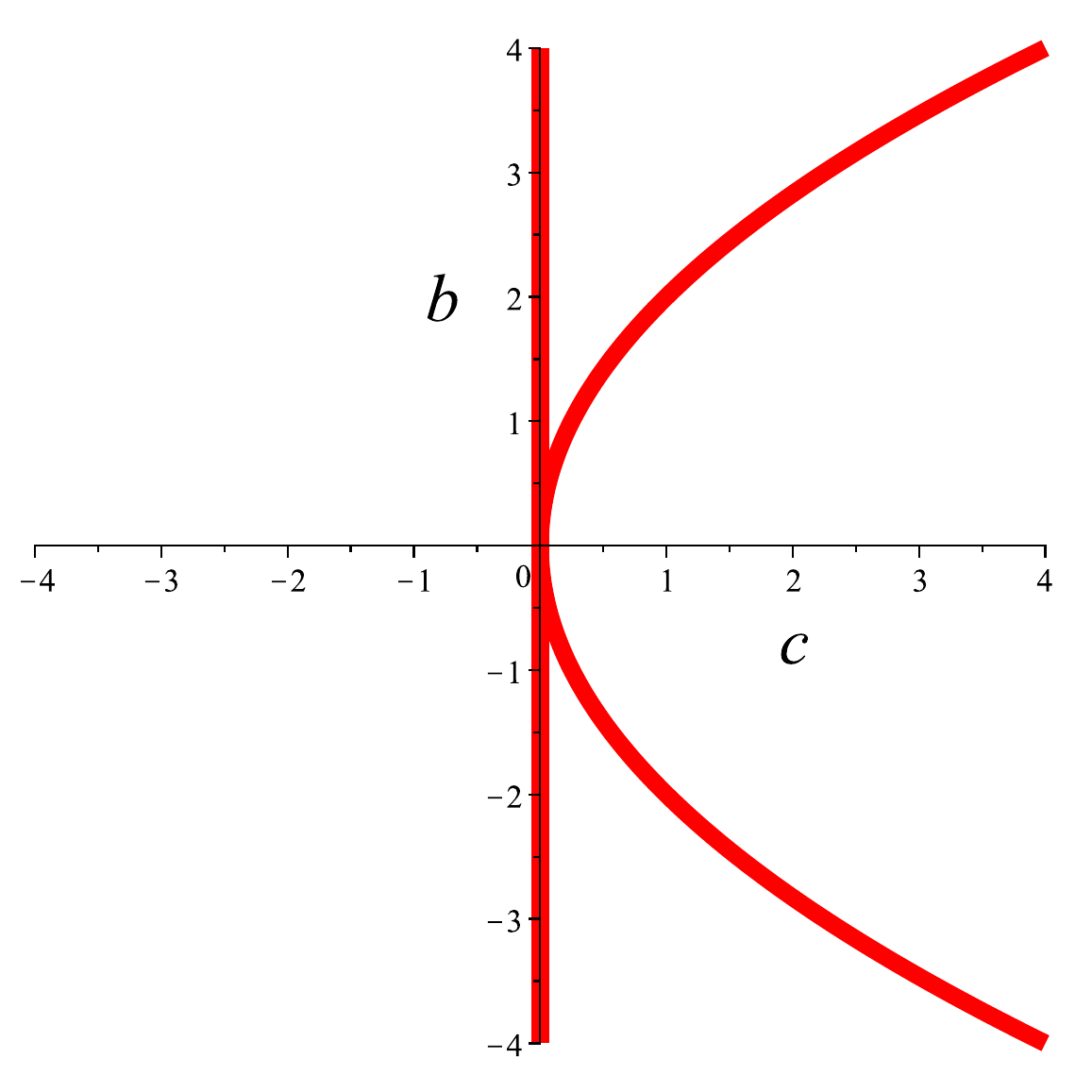}}
	\subfloat[]{\label{Fig:Degree_2_equation_open_CAD_positive}\includegraphics[width=4.5cm]{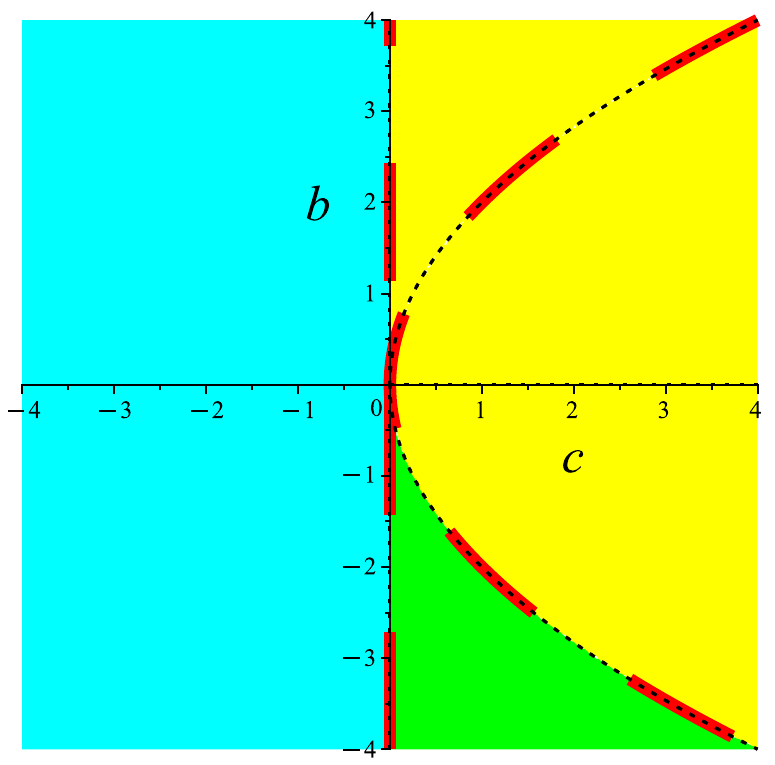}}
	\caption{(a) The discriminant variety of the parametric system $x^2+bx+c=0$ when one studies the number of \emph{positive} real solutions. The open CAD of the parameter plane with respect to this is the same as \Cref{Fig:Degree_2_equation_open_CAD}. (b) The system has one, two and no positive solutions in the cyan, green and yellow coloured regions respectively.}
	\label{Fig:Degree_2_equation_r}
\end{figure}

So in this case the discriminant variety is $V(4c-b^2) \cup V(c)$, as shown in \Cref{Fig:Degree_2_equation_discriminant_positive}. The open CAD gives the same four cells as the previous case. However, previously the line $c=0$ was computed as part of the CAD process while this time it was an explicit input to the system.  Testing the sample points we find the system has one positive solution over Cell 1, two positive solutions over Cell 2, and no positive solutions over Cells 3 and 4:  as visualised in \Cref{Fig:Degree_2_equation_open_CAD_positive}.  For a more involved example see the work of \cite{Lichtblau-2021} which applied the approach to a problem from chemical reaction network theory. 

\subsection{Discriminant variety and elimination}
\label{sec:Discrim_and_elimination_problem}

Before we continue there is a need to clarify of several terminologies to avoid confusion. A set of equations and inequalities of parametric polynomials from $R=\mathbb{Q}[k_1,\dots,k_r][x_1,\dots,x_n]$ is simply called a \emph{system}. A non-zero polynomial $f\in\mathbb{Q}[k_1,\dots,k_r]$ whose variety contains the boundaries between different parameter regions where number of real solutions of a given system varies, is called a \emph{discriminant polynomial} for this system and $V(f)$ is called a \emph{discriminant variety}. The \emph{minimal discriminant variety} of a system is the smallest discriminant variety of the system. As we observed in the simple example of Section~\ref{sec:Restricting_to_positive_solutions}, a discriminant variety can have more than one component corresponding to different behaviours of the system. In this case $b^2-4c$ for multiplicity of solutions (singular locus), and $c$ for sign of solutions, the (minimal) discriminant polynomial is the product of these two components, $c(b^2-4c)$. See \cite{Lazard-Rouillier-2007} for complete list of components of the minimal discriminant variety. In this work we use loosely the terminology of discriminant variety for a single component as well, for example the component coming from multiplicity of solutions.

As noticed in the Subsections~\ref{sec:Method_and_example} and \ref{sec:Restricting_to_positive_solutions}, to solve the problem stated in Section~\ref{sec:Problem_Statement}, which we refer to as \textbf{Problem 1} from here on, one needs to compute a discriminant variety and for that, several elimination problems of the following form, being referred as \emph{Problem 2} need to be solved.\smallskip 

\textbf{Problem 2:} Let $F'=\lbrace f_1,\dots,f_{m'}\rbrace\subset \mathbb{Q}[y_1,\dots,y_r,y_{r+1},\dots,y_{r+n}]$. Find a polynomial $h\in\mathbb{Q}[y_1,\dots,y_r]$ such that $\pi_{r}^{r+n}\big(V(F')\big)\subseteq V(h)$.\smallskip

A trivial solution to this problem is always $h=0$. But when $\pi_{r}^{r+n}\big(V(F')\big)\neq \mathbb{R}^n$, one is interested in a non-trivial $h$ and preferably of the lowest possible degree. Note that the Gr\"obner basis method mentioned in Section~\ref{sec:Method_and_example} does not necessarily provide the solution to this problem with the minimum degree. It can be the case that the Zariski closure of the projection is not generated by a principal ideal, or some of the components of the projection are purely complex with no real point in them, for example if the GB method returns a polynomial such as $x^2+1$ which has no real solution: in that case the minimum degree solution to the problem is $h=1$ rather than $x^2+1$.

The component of a discriminant variety for Problem 1, when $m=n$, coming from the singular locus can be computed by Problem 2 after letting $y_i=k_i$, $i=1,\dots,r$, $y_{r+i}=x_i$, $i=1,\dots,n$, $m'=n+1$ and $f_{n+1}=d$ where $d$ is determinant of the Jacobian matrix of $F=\lbrace f_1,\dots,f_n\rbrace$ with respect to $\lbrace x_1,\dots,x_n\rbrace$.\footnote{See \cite[Section 2.2]{Cutkosky-2004} for what to pick instead of $d$ when $m\neq n$ or $F\mid_{k=k^\star}$ is not generically 0-dimensional.}

\section{Recent Prior work on our Population Model Application}
\label{sec:Recent_Prior_work_on_Population_Model_Application}

Recently, in \cite{Rost-Sadeghimanesh-2021-1,Rost-Sadeghimanesh-2021-2}, the problem described in \Cref{sec:Population_Models_with_the_Allee_Effect} was studied using, amongst others, the tools just introduced in \Cref{sec:Solution_via_CAD_and_Discriminant_Variety}.    
The combination of discriminant variety and CAD implemented in Maple's \texttt{RootFinding:-Parametric} package was used on a normal laptop to successfully study two populations with the strong Allee effect in \cite{Rost-Sadeghimanesh-2021-2}. However, when attempted for the three population case in \cite{Rost-Sadeghimanesh-2021-1} this algorithm did not terminate on a normal laptop before running into memory limitations.  

An investigation into the problem identified that the first step of the algorithm, the computation of the discriminant variety, was infeasible on its own before any CAD was produced. The discriminant variety is usually computed via GB techniques\footnote{Details on the \texttt{DiscriminantVariety} command from \texttt{RootFinding:-Parametric} in Maple:  \url{https://www.maplesoft.com/support/help/Maple/view.aspx?path=RootFinding\%2fParametric\%2fDiscriminantVariety}.}.  
Unfortunately, in this case the GB computation needed to compute the discriminant variety is not feasible on a normal computer (e.g. the one specified in Section \ref{sec:Computational_environment}).  
To overcome this problem, in \cite{Rost-Sadeghimanesh-2021-1}, a combination of this algebraic method, CAD with respect to the discriminant variety, with a numerical sampling approach was developed, to build an approximation of the requested decomposition of the parameter space.  We note a similar numeric-algebraic approach was used for studying chemical reaction networks in the work of \cite{EEGRSW17}.

\begin{figure}[ht]
	\centering
	\subfloat[]{\label{Fig:numeric_algebraic_step_1}\includegraphics[width=4cm]{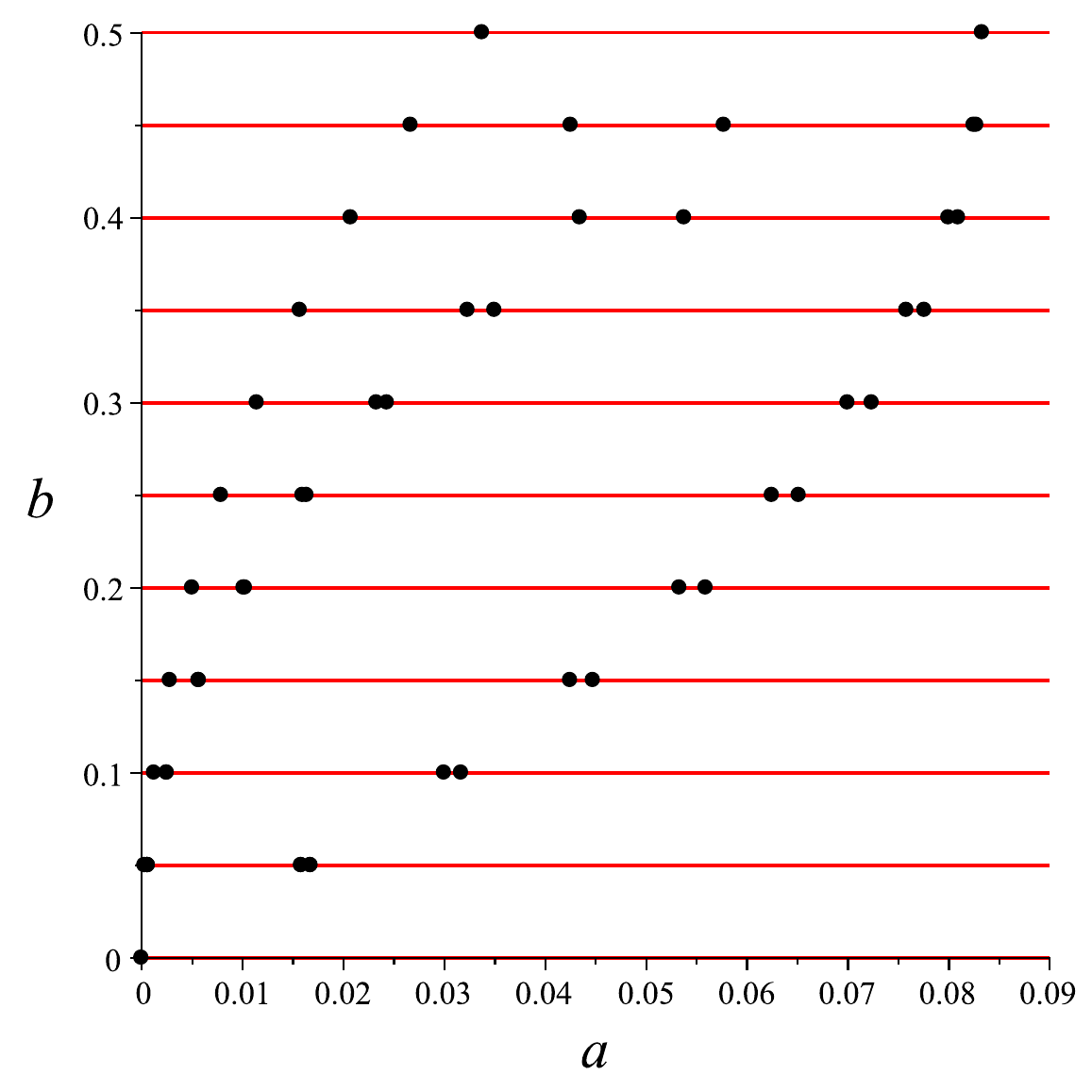}}
	\subfloat[]{\label{Fig:numeric_algebraic_step_2_1}\includegraphics[width=4cm]{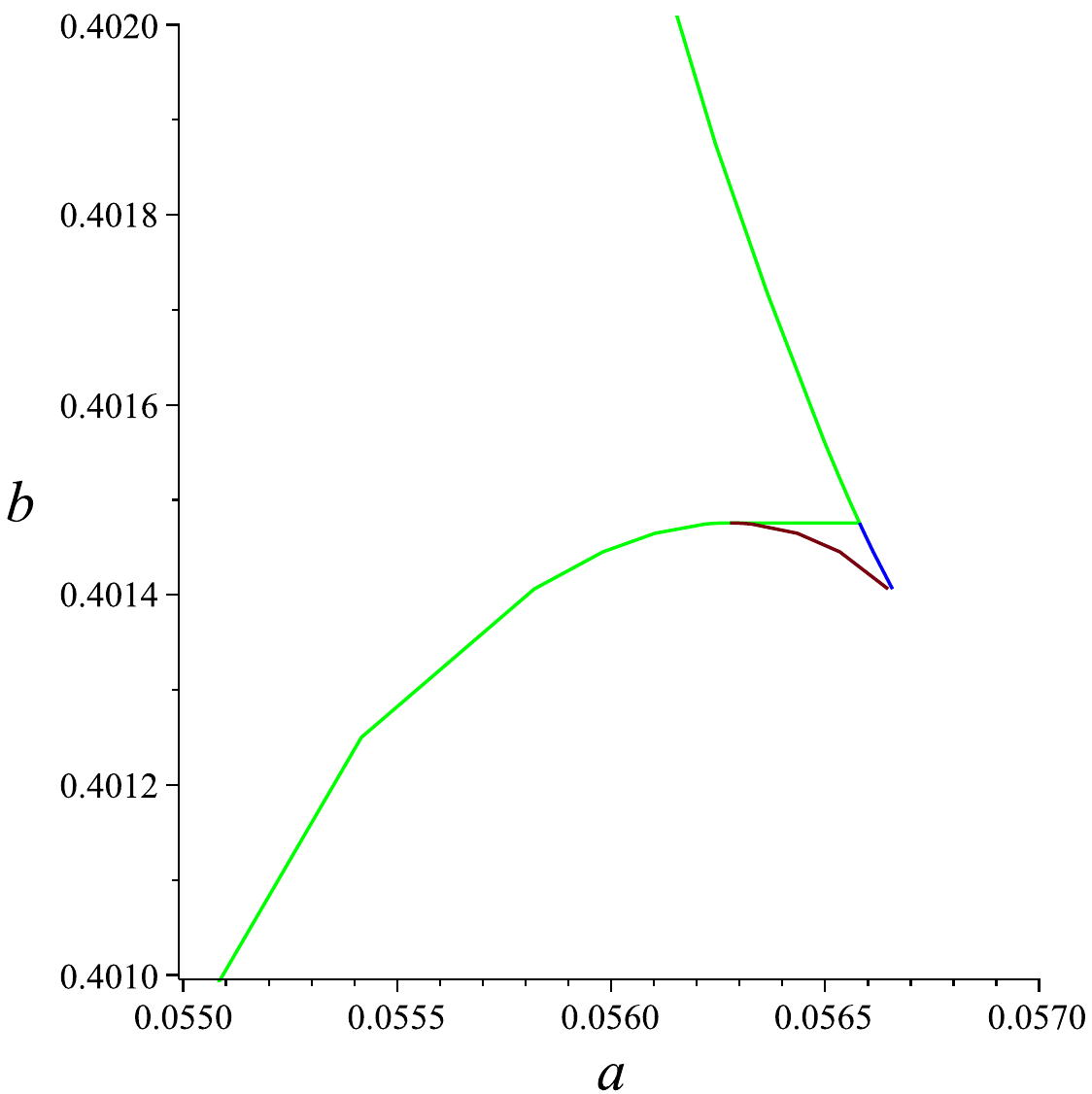}}
	\subfloat[]{\label{Fig:numeric_algebraic_step_2_2}\includegraphics[width=4cm]{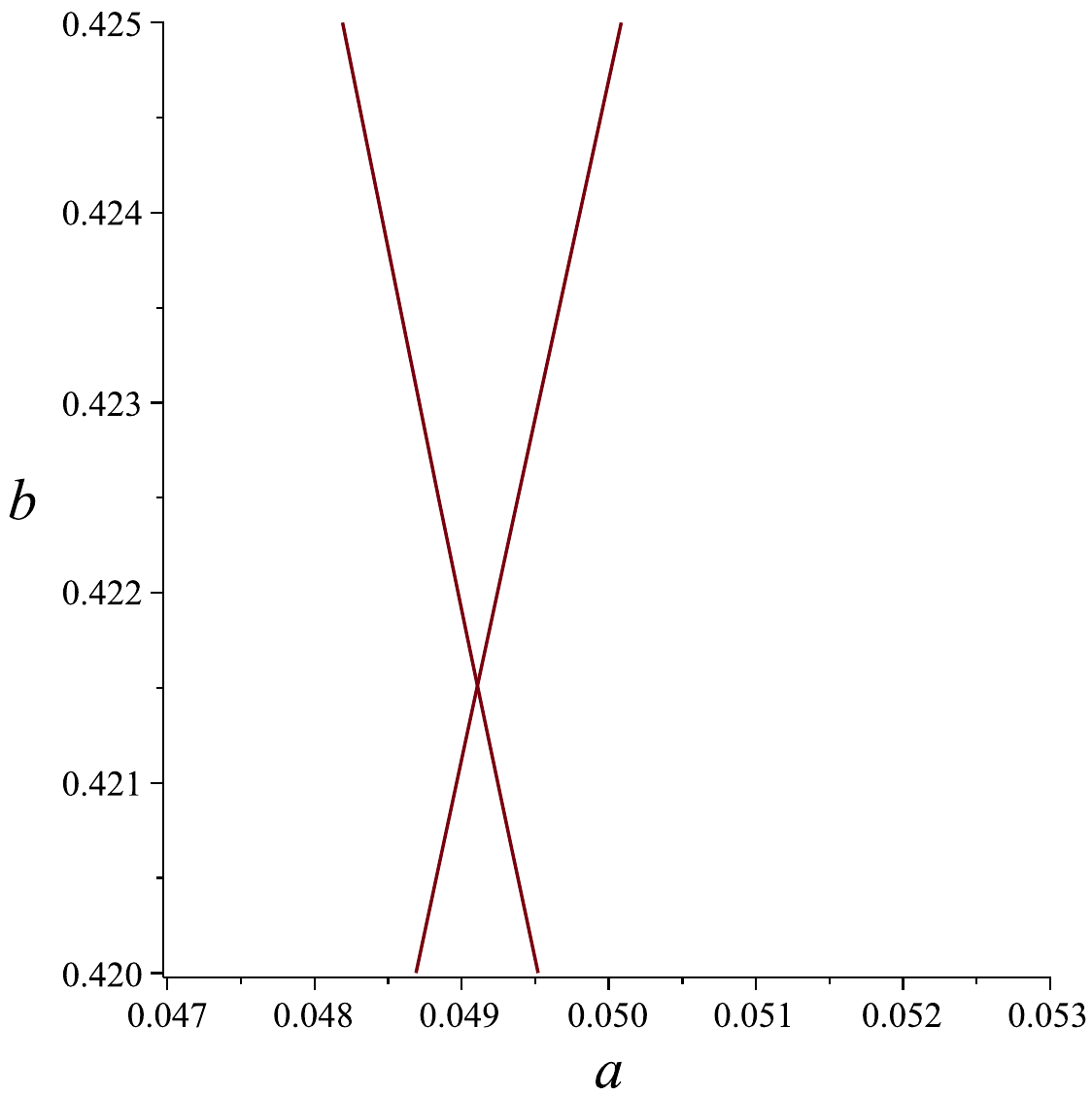}}\\
	\subfloat[]{\label{Fig:numeric_algebraic_result}\includegraphics[width=4cm]{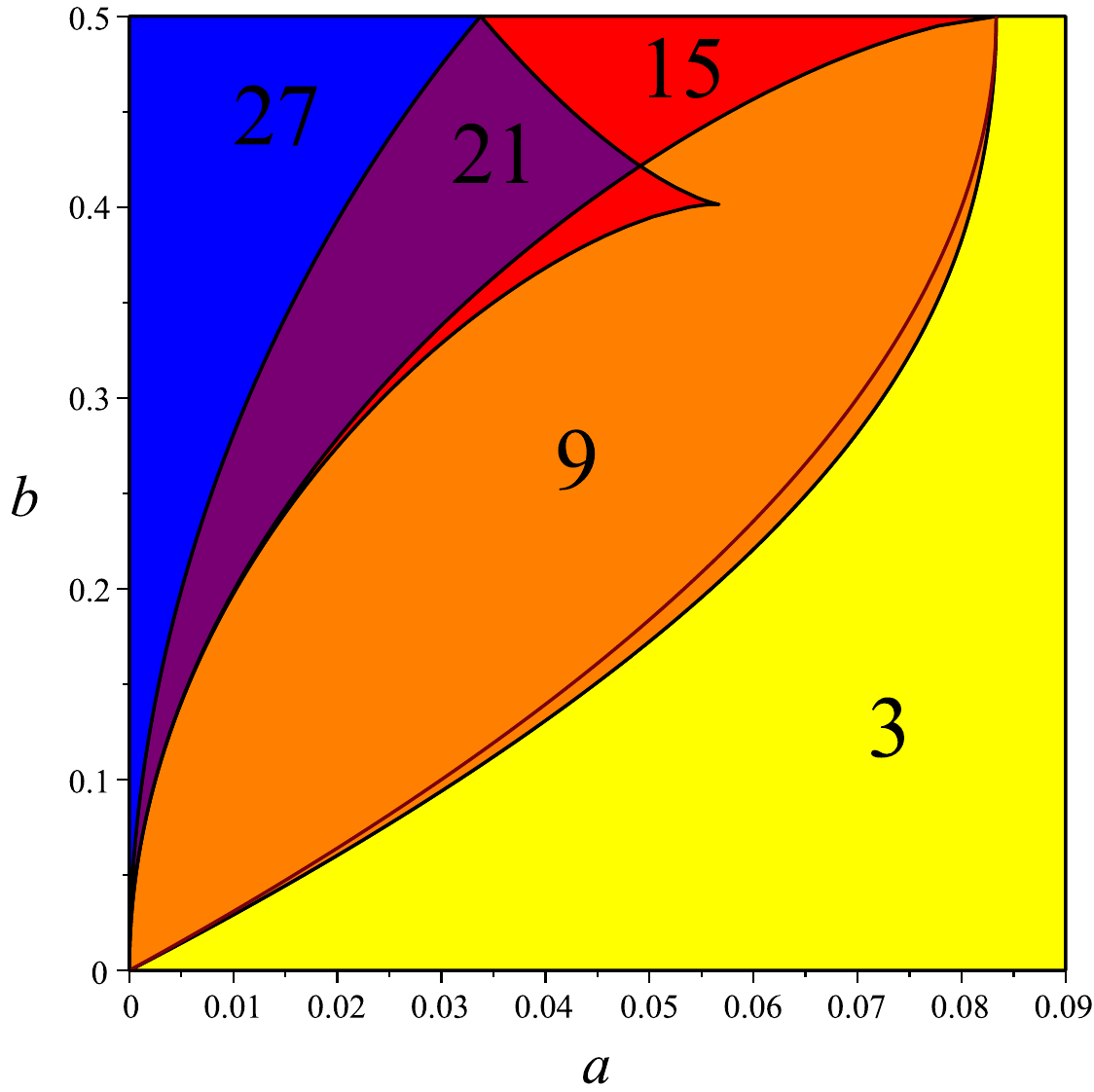}}
	\subfloat[]{\label{Fig:numeric_algebraic_result_zoomed}\includegraphics[width=4cm]{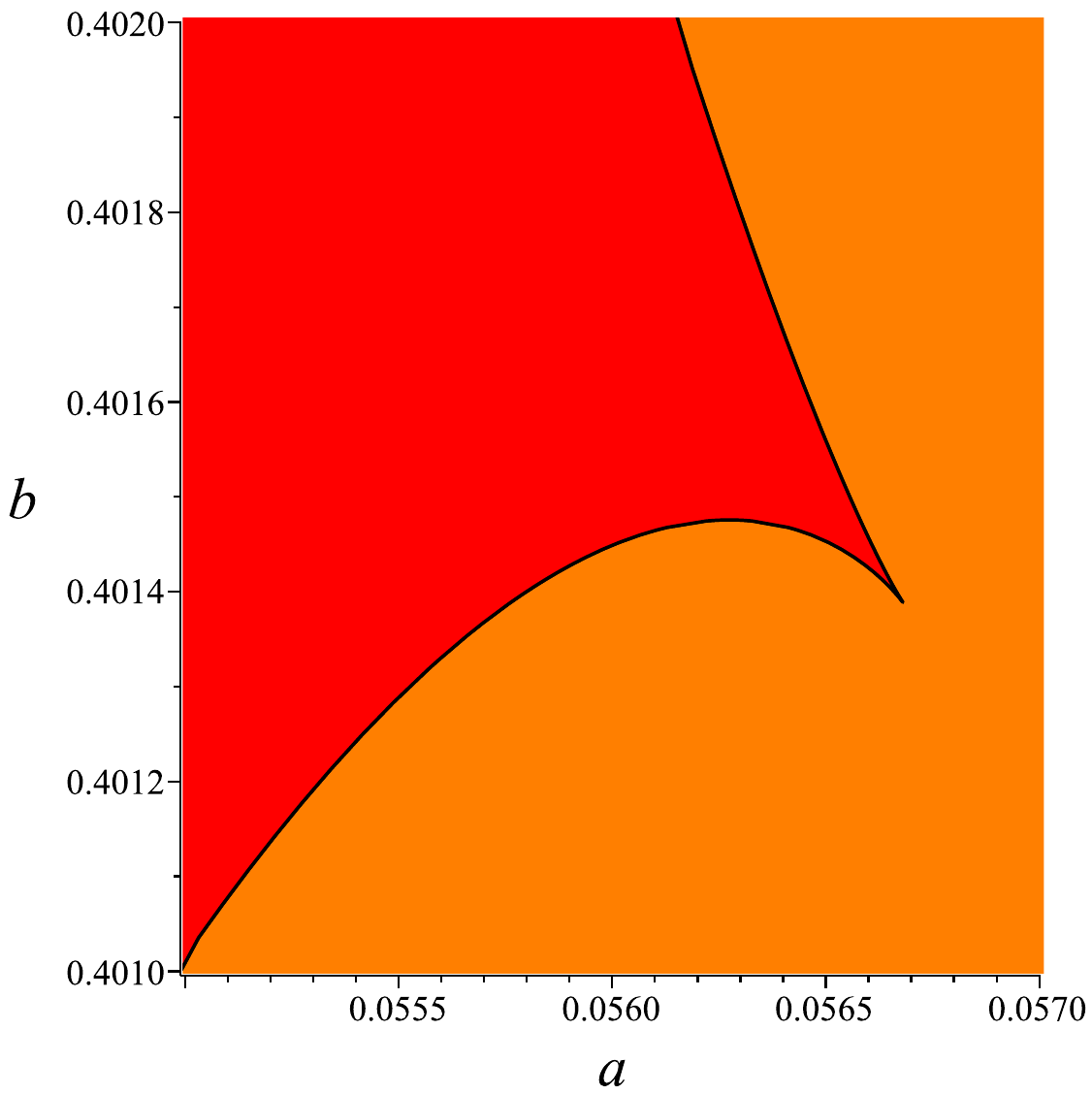}}
	\caption{\Cref{Fig:numeric_algebraic_step_2_1,Fig:numeric_algebraic_step_2_2,Fig:numeric_algebraic_result,Fig:numeric_algebraic_result_zoomed} are taken from \cite[Figures 2 and 3]{Rost-Sadeghimanesh-2021-1}. They demonstrate the idea behind the numeric-algebraic algorithm developed in \cite{Rost-Sadeghimanesh-2021-1} and the result of its implementation on the three populations with strong Allee effect example. (a) The first step is to find the intersection of the discriminant variety with several hyperplanes defined by fixing a value of one of the parameters, in this case the red colour horizontal lines. (b-c) The second step is to run a numeric search to find where the number of intersection points in the previous step change. (d) The final output of the algorithm for the three populations example. (e) Enlargement of a tiny region from part (d) related to the intersection identified in (b).}
	\label{Fig:numeric_algebraic_}
\end{figure}

The numeric-algebraic algorithm in \cite{Rost-Sadeghimanesh-2021-1} consists of two steps. The first step is to fix a value of one of the parameters and use the algebraic algorithm with one parameter less. This gives the intersection of the discriminant variety with the hyperplanes defined by the fixing of the other parameter. \Cref{Fig:numeric_algebraic_step_1} shows the result of this step for 11 equally distanced values of $b$ between and including 0 and 1/2. The next step is using a numeric search and again the algebraic algorithm with one parameter fixed, to find regions where the behaviour of the intersection of the discriminant variety with the horizontal lines changes. This step finds two regions \cite[Figure 2]{Rost-Sadeghimanesh-2021-1} shown in \Cref{Fig:numeric_algebraic_step_2_1} and \Cref{Fig:numeric_algebraic_step_2_2}. The final output of this algorithm is \cite[Figure 3]{Rost-Sadeghimanesh-2021-1} visualised in \Cref{Fig:numeric_algebraic_result}.  This approach can guarantee the behaviour of the system up to precision chosen in the numeric-algebraic algorithm, in this case 7 digits after the decimal point. 

The method does not guarantee that there is no smaller region with different behaviour that may have been missed. For increased confidence the user could re-run the algorithm with a higher precision requested.  However, it is not possible to achieve full certainty, and so a symbolic verification of the results is still desired.

\Cref{Fig:numeric_algebraic_result_zoomed} is a magnified portion of \Cref{Fig:numeric_algebraic_result} and shows a surprising finding identified by this  analysis:  it was expected that when increasing the dispersal rate, $a$ the number of steady states of the network should monotonically decrease, however, the zoomed-in region shows there is a possibility to temporarily increase the number of steady states for some choices of the Allee threshold, $b$.

\section{New Approaches using Resultants}
\label{sec:New_Approach_using_Resultants}

We now report on some new, purely algebraic, approaches to our problem, which prove sufficient to tackle the three population case.  They do not use GB to compute the discriminant variety, but instead use the theory of resultants.  The ideas were motivated by the improved CAD performance available when the input contains equational constraints \cite{EBD20}, but are developed here outside of the CAD context allowing for simplicity of presentation and greater savings.

\subsection{Using a single univariate resultant}\label{sec:Using_Simple_Resultant}

We start by considering the simplest use of a resultant. Recall \textbf{Problem 2}. In the case where $n=1$ and $m'=2$, an alternative to a Gr\"obner basis for the elimination of variable $y_{r+1}$, denoted by $x$ here, to get polynomials only involving parameters is to use a resultant. 

We denote the resultant of $f_1$ and $f_2$ with respect to $x$ by $g=\res(f_1,f_2,x)$.  The resultant is a polynomial expression in which $x$ has been eliminated, and which is equal to zero if and only if the polynomials $f_1$ and $f_2$ have a common root.  It may be calculated as the determinant of the Sylvester Matrix of the two polynomials (a square matrix of size the sum of the degrees in $x$ of the polynomials, formed from the coefficients of the powers of $x$).

Note that when used for \textbf{Problem 1}, $V(g)$ contains the discriminant variety, but also possibly more components. Containment of the discriminant variety is enough to ensure that an open CAD of the parameter space with respect to $V(g)$ as in \Cref{sec:Solution_via_CAD_and_Discriminant_Variety} solves our problem. 

For the simple example in \Cref{sec:Method_and_example} we have one variable $x$ and two polynomials $f$ and $f'$.  The resultant of these two polynomials with respect to $x$ is
\[\res(f,f',x)=\begin{vmatrix}
	1 & 2 & 0\\
	b & b & 2\\
	c & 0 & b
\end{vmatrix}=-(b^2-4c).\]
I.e. the same polynomial that we found by the Gr\"obner basis, up to a sign. For the extra part of the discriminant variety of \Cref{sec:Restricting_to_positive_solutions}, again we have one variable $x$ and two polynomials $f$ and $x$, the corresponding resultant is the following.
\[\res(f,x,x)=\begin{vmatrix}
	1 & 1 & 0\\
	b & 0 & 1\\
	c & 0 & 0
\end{vmatrix}=c.\]
This is again the same polynomial computed by the Gr\"obner basis approach. Therefore for this simple example, replacing the Gr\"obner basis computation with resultants did not change the discriminant variety obtained.

The question now is what about the case of having more than two polynomials and more than one variable? There are several generalisations of resultants to multi-polynomial cases such as the multi-polynomial resultant for $n$ polynomials in $n$ variables using homogenisation, or Waerden's $u$-resultant (see \cite[Chapter 3]{Cox_Using_Algebraic_Geometry} for both), or the Dixon resultant for $n+1$ polynomials in $n$ variables, introduced next. In the applications of our interest, such as population dynamics and Chemical Reaction Network theory, we always have $n$ equations in $n$ variables and an extra polynomial coming from determinant of the Jacobian matrix or positivity constraints etc. Therefore the Dixon resultant is a suitable generalisation to try.

\subsection{Using the multivariate Dixon resultant}\label{sec:Using_Dixon_resultant}

The Dixon resultant is named after Arthur Lee Dixon, a British mathematician who extended the B\'{e}zout-Cayley method of computing the simple resultant for computing resultant of three polynomials in two variables \cite{Chtcherba-Kapur-Minimair-2005,Dixon-1909,Dixon-1909-2}.  It uses only one determinant to produce a polynomial only involving parameters for a system of $n+1$ equations in $n$ variables that vanishes whenever the system has a solution. Therefore if we denote the Dixon resultant of the system $F$ of the problem statement in \Cref{sec:Problem_Statement} by $g$, then $V(g)$ contains the discriminant variety. For a simple explanation of all the steps of computing the Dixon resultant see \cite[Section 2.1]{Minimair_DR_package}. Here we do not explain the general algorithm, but only show the computations for the simple example of \Cref{sec:Method_and_example}.

Define an auxiliary variable $\bar{x}$ and consider the following matrix:
\[M_1=\begin{bmatrix}
	x^2+bx+c & 2x+b\\
	\bar{x}^2+b\bar{x}+c & 2\bar{x}+b
\end{bmatrix}.\]
The Dixon polynomial is defined as $\det(M_1)/(x-\bar{x})$, denoting it by $p$, we have $p=2x\bar{x}+b(x+\bar{x})+b^2-2c$. Then we have to solve the linear system of equations created by considering:
\[\begin{bmatrix}
	\bar{x} & 1
\end{bmatrix}\begin{bmatrix}
	a_{11} & a_{12}\\
	a_{21} & a_{22}
\end{bmatrix}\begin{bmatrix}
	x\\
	1
\end{bmatrix}=p.\]
Solving gives us the Dixon matrix,
\[M_2=\begin{bmatrix}
	2 & b\\
	b & b^2-2c
\end{bmatrix}.\]
Now the Dixon resultant is the determinant of a maximal-rank submatrix of $M_2$ which is $M_2$ itself. We get $\det(M_2)=b^2-4c$ as expected.

So the first part of the algorithm in \Cref{sec:Solution_via_CAD_and_Discriminant_Variety} may be replaced by the Dixon resultant. There are several implementations of the Dixon resultant, for example based on how to find the maximal-rank submatrix or how to compute the determinants. One implementation can be found in the computer algebra system Fermat\footnote{\url{https://home.bway.net/lewis}} \cite{Lewis-2004}, and another implementation is within a Maple package called DR\footnote{\url{https://github.com/mincode/dixon}} \cite{Minimair_DR_package}.

A complexity analysis on construction of the Dixon resultant matrix is given in \cite[Theorem 3.1]{Qin_Dixon_res_complexity} and is of order of $\mathcal{O}(n!^3\mathfrak{m}^{4n})$ where $\mathfrak{m}$ is the maximal univariate degree of the polynomials in $F$ in each of its variables. This complexity is somewhere between singly exponential and doubly exponential; lower than the worst case complexity of Gr\"obner basis computation which is doubly exponential in $n+r$ (with the Gr\"obner basis computation is in the ring $\mathbb{Q}[k_1,\dots,k_r,x_1,\dots,x_n]$) \cite{Mayr-Meyer-1982,Mayr-Ritscher-2013}. We remark that there is a singly exponential algorithm to compute Gr\"obner bases and thus the elimination task of our interest and the computation of discriminant variety \cite{Guillaume-2006,Guillaume-2011}. However the size of matrices involved in the computation are often so big that makes it not practical and therefore the current implementations of the GB based methods are still with worst case complexity doubly exponential.

\subsection{Using a chain of univariate resultants}\label{sec:Using_resultant_chain}

We next consider an alternative to the Dixon resultant based on iterated use of the  univariate resultant from \Cref{sec:Using_Simple_Resultant}.

Let $f_1$ and $f_2$ be two polynomials in $N$ variables, $x_1,\dots,x_N$. Then by Theorem 8 of \cite[Chapter 3]{Cox_Ideals_Varieties_and_Algorithms}, if the degrees of $f_1$ and $f_2$ in $x_N$ are positive, we have that
\[
\pi_{N-1}^N\big(V(f_1,f_2)\big)\subseteq V\big(\res(f_1,f_2,x_N)\big).
\]
In other words, $V(\res(f_1,f_2,x_N))$ contains the projection of the set of common solutions of $f_1$ and $f_2$ into their first $N-1$ coordinates. On the other hand, if the degree of either of these two polynomials in $x_N$ is not positive, then $\pi_{N-1}^N\big(V(f_1,f_2)\big)$ is a subset of the solution set of the one (or both) which does not involve $x_N$ as polynomials in $N-1$ variables. 

Whenever both polynomials involve $x_N$, we follow the standard definition of resultant as Sylvester determinant, otherwise we redefine the resultant to be 1 if at least one of the two polynomials is a non-zero constant; and if not, then in the cases where the degree in $x_N$ of $f_1$, or $f_2$, or both is not positive, we will redefine $\res(f_1,f_2,x_N)$ to be respectively $f_1$, $f_2$ or $f_1f_2$.  Note this gives different results to the Sylvester determinant (which would be a power of the polynomial for the first two cases and a constant for the latter). 

\begin{remark}\label{remark:resultant_both_polys_not_having_xN}
	A better choice for the case where both $f_1$ and $f_2$ do not involve $x_N$ is $f_1^2+f_2^2$, whose variety over the reals is equal to $V(f_1,f_2)=V(f_1)\cap V(f_2)$.  However, to control the growth of the size of the polynomials in the algorithms coming later, we opted for $f_1f_2$ whose variety $V(f_1)\cup V(f_2)$ is a superset of the intersection.
\end{remark}

Assume $F$ contains $m$ polynomials in $N$ variables, where $m>N$. We now have that
\begin{align*}
	\pi_{N-1}^N\big(V(F)\big) & = \bigcap_{i=2}^m \pi_{N-1}^N\big((V(f_1,f_i)\big)\\
	& \subseteq  \bigcap_{i=2}^m V(\res(f_1,f_i,x_N)).
\end{align*}
Taking this set of resultants gives us $m-1$ polynomials in $N-1$ variables. 

This gives us a route to use iterated univariate resultants to solve our problem. We refer to this technique as \emph{res-chain} standing for a \emph{chain of univariate resultants}. Let us return to the parametric polynomial ring in \Cref{sec:Problem_Statement}, with $n$ variables and $r$ parameters. Let $m$ be the number of polynomials in the original system of equations plus the extra polynomials needed to study the discriminant variety. If $m > n$, then repeating the above process iteratively over $n$ steps, gives one $m-n$ polynomials involving only parameters. We denote these polynomials by $g_i$, $i=1,\dots,m-n$.  Then $V(g_1,\dots,g_{m-n})$ contains the discriminant variety.

\subsection{Degree drops and constant evaluations}

Note that when taking the resultant $\res(f_1, f_2, x_i)$ it is possible that the polynomial produced does not have positive degree in $x_{i-1}$.  In this case, the resultants taken in the subsequent stage will be evaluated according to the modified resultant definition above.  This has the effect of passing the information down to the relevant level.

Let us now consider what happens when a resultant evaluates to zero.  This means that the two input polynomials have a common root everywhere.  This would happen if the two polynomials have a common factor.  If in the above process all $m$ polynomials share a common factor we can return $0$ as the defining polynomial of the projection. But if only some of them share a common factor, then we will continue the process in two separate branches.  We explain this by means of a simple example system with $f_1=h_0h_1$ and $f_2=h_0h_2$ where $h_1$ and $h_2$ are relatively prime and $h_0$ is not a factor of $f_i$ for $i=3,\dots,m$. We have that
\begin{align*}
	\pi_{N-1}^N\big(V(F)\big) &= \pi_{N-1}^N\big(V(h_1, h_2, f_3, \dots, f_m) \cup V(h_0, f_3, \dots, f_m)\big)\\
	&= \pi_{N-1}^N\big(V(h_1, h_2, f_3, \dots, f_m)\big) \bigcup \pi_{N-1}^N\big(V(h_0, f_3, \dots, f_m)\big).
\end{align*}

Thus one can apply the former process on these two branches and at the end take the union of the final output sets of polynomials where the variables are eliminated.

If a resultant evaluates to a non-zero constant then it means that the two input polynomials can never share a common root.  Should this happen it means there is no solution to the problem and the algorithm can terminate.  Although if we have branched as above then we could only terminate that branch.  

\subsection{Efficiencies from factorisation}

Using the fact that 
\[
\res(p_1p_2,p_3,x)=\res(p_1,p_3,x)\res(p_2,p_3,x),
\]
one can modify the above process so that the size of the Sylvester Matrix determinants needed to be computed gets smaller. Suppose we are computing the resultant of two polynomials $p=\prod_{i=1}^s p_i^{\alpha_i}$ and $q=\prod_{i=1}^t q_i^{\beta_i}$, where $p_i$ and $q_j$ are irreducible factors of $p$ and $q$.  Then instead of a single determinant of size $\sum_{i=1}^s\alpha_i\degree(p_i)+\sum_{i=1}^t\beta_i\degree(q_i)$ one can use $s \times t$ determinants of sizes $\degree(p_i)+\degree(q_j)$.

This should mean less computation resources. For example, consider the simplified situation where $\hat{p}$ is the maximum degree of any factor, $\hat{s}$ is the maximum number of factors and there are no repeated factors (i.e. $\alpha_i=\beta_i=1$ for all $i$).  Then we are comparing a single determinant of size $2\hat{s}\hat{p}$ with $\hat{s}^2$ determinants of size $2\hat{p}$.  Calculating the determinant with cost $\mathcal{O}(n^3)$ for matrix of size $n$ means we save a factor of $\hat{s}$.  Such a saving is repeated at each stage in our chain of univariate resultant computation.

\subsection{Algorithm with a simple chain of resultants}

Following this analysis we implemented an algorithm in Maple called \texttt{ResChainSimple} (\Cref{Alg:ResChainSimple}).  The algorithm takes a list of polynomials and eliminates variables to produce a set of polynomials; the union of whose varieties has the projection as a subset.

Within the algorithm, the master list $L$ contains sublists, denoted $B_k$ for each of the different branches in the analysis:  we start with only one branch in the initialisation.  Each branch contains a set of polynomials, each represented by a list, $P_j$ containing their irreducible factors.  
The outer loop (the for-loop on $i$) concerns the levels of projection.  At each level the while-loop processes one branch at a time (noting that an iteration of this loop can create further branches in $L$ to process).  

So long as a branch contains more than one polynomial with a variable we enter the for-loop on $j$ which takes the resultants of the factors of the first polynomial with the others.  Should a common factor be found then an additional branch is created, as discussed earlier.

If a branch contains only one polynomial (represented as a list of factors) involving a variable, then the algorithm terminates and returns a single polynomial, $0$.  This trivially meets the specification (in the sense that $V(0)$ is the whole space and thus has the projection as a subset) but of course this is not useful. 
This output indicates that the algorithm cannot use resultants to eliminate all the variables, and the user would be advised to seek a different approach.  We note that triggering this case terminates not just the branch but the whole algorithm, continuing the other branches would be pointless since the union of their output with $V(0)$ would simply be $V(0)$.  We expect that progress can be made on such cases in future work.

In the implementation of Algorithm~\ref{Alg:ResChainSimple}, we included a check point for lines $19-22$ to avoid adding a new branch more than once. This is not critical for correctness of the algorithm, but if not considered, can slow down the algorithm by spending time on processing repeated branches. As an example of such a case consider a current branch under process containing the polynomials $f_1f_2$, $f_1g$, $f_1h$, $f_1p$ and $q$ where $f_2$, $g$, $h$, $p$ and $q$ are relatively prime with respect to $f_1$. In this case the for-loop of line 16 will add the new branch $(f_1,q)$ three times.

The implementation also avoids adding a branch $(P_1',\dots,P_{m-1}')$ to $L'$ (line 25) if at least one of the polynomials corresponding to $P_i'$ is a constant.

\begin{algorithm}[ht!]
	\caption{Simple approach to use iterated univariate resultants to eliminate variables from equations.}
	\label{Alg:ResChainSimple}
	\begin{algorithmic}[1]
		\Procedure{\texttt{ResChainSimple}}{$F$, $x=(x_1,\dots,x_n)$}
		\Inputo $F$ is a list of $m$ polynomials and $x$ is a tuple of $n$ variables.  The polynomials are defined in the $n$ variables and an additional $r$ parameters.
		\Outputo $S$, a set of irreducible polynomials none of which involve the variables in $x$. The projection of $V(F)$ as a subset of $\mathbb{R}^{r+n}$ into $\mathbb{R}^r$, is a subset of $\cup_{p\in S}V(p)$.
		\State \emph{Initialisation:} Let $S=\lbrace\;\rbrace$. Denote the set of irreducible factors of the $i$th element of $F$ with $P_i$ for $i=1,\dots,m$. Let $B_1=(P_1,\dots,P_m)$, and $L=(B_1)$. I.e. there is only one branch at the start of the algorithm.
		\For{$i$ \textbf{from} $n$ \textbf{by} $-1$ \textbf{to} $1$}
		\State Let $L'=(\;)$
		\While{$L\neq (\;)$}
		\State Pick a branch from $L$, denote it by $B$.
		\State Remove $B$ from $L$.
		\If{$B$ contains only one list}
		\State Let $P$ denote the only member of $B$.
		\If{any polynomial in $P$ contains any of $x_1,\dots,x_i$}
		\State \Return $\lbrace 0\rbrace$.
		\Else
		\State $S=S\cup P$.
		\EndIf
		\Else
		\State Let $B = (P_1, \dots, P_m)$.
		\For{$j$ \textbf{from} $1$ \textbf{by} $1$ \textbf{to} $m-1$}
		\State $P_j'=(\;)$.
		\For{$f$ \textbf{in} $P_1$}
		\For{$g$ \textbf{in} $P_{j+1}$}
		\If{$f=g$}
		\State Set $B'$ to the list found by removing all $P_\ell$ with $f$ from $B$, and prepend a list $(f)$ to its start.
		\State Append $B'$ to $L$.
		\Else
		\State Append irreducible factors of $\res(f,g,x_i)$ to $P_j'$.
		\EndIf
		\EndFor
		\EndFor
		\EndFor
		\State Append $(P_1',\dots,P_{m-1}')$ to $L'$.
		\EndIf
		\EndWhile
		\State Replace $L$ with $L'$.
		\EndFor
		\State Update $S$ with its union with the sets in all branches in $L$.
		\State \Return $S$.
		\EndProcedure
	\end{algorithmic}
\end{algorithm}

\subsubsection{Case analysis}

Consider the case where there is no branching because of polynomials sharing a common factor.  First suppose that $m=n+1$. Then we only have one branch in each step and the factors of the $j$th polynomial in any step of the iteration are resultants of the factors of the first polynomial and factors of the $(j+1)$-th polynomial in the step before. Finally the last step has only one polynomial and its solution set is the union of solution sets of its factors.  

Next suppose $m>n+1$.  Then in the last step we have $m-n$ polynomials and the projection is the intersection of the solution set of each of these $m-n$ polynomials. But this intersection set is still a subset of the union of solution sets of factors of all of the $m-n$ polynomials, and as the final output the algorithm just returns the set of all factors of the polynomials in the last step.

In the case where $m<n$, the algorithm will most likely end up with a branch with a single polynomial involving a variable, thus returning $0$. However, this is not a certainty: it might be that some variables are not present, or they automatically get eliminated in the projection steps of the other variables, so that the algorithm ends up finding the projection even though we have fewer polynomials than variables.

If branching due to common factors occurs then we can no longer conclude a successful output just because $m \geq n+1$.  The branches created have fewer polynomials, which increases the likelihood of hitting the case where we return $0$.  If all branches avoid this case then the final output is the union of the output for each branch.

\subsubsection{Algorithm sensitivities}

Note that this algorithm is sensitive to the order of the variables in $x$ (and likewise the open CAD to follow will be sensitive to the order of the parameters).  We expect this choice of order may have a significant effect, as is already well known for CAD\footnote{For CAD there is a theoretical analysis of the impact of the ordering in \cite{BD07}, numerous optimisation attempts e.g. \cite{DSS04}, and more recently experimentation with machine learning technology to make the choice \cite{dRE24,FE20b,HEWBDP19}.}.  

Further, \Cref{Alg:ResChainSimple} is sensitive to the order of the polynomials which appear in $F$:  it is clear from the algorithm that the first polynomial is treated specially, but note that the order of subsequent polynomials will affect which polynomials are positioned first in subsequent levels, and so has an effect also.  Heuristic choices for both of these orderings are a potential topic for further study.

\subsection{Algorithm with branching resultant chains}

It should not be surprising that \Cref{Alg:ResChainSimple} can potentially generate many extra components that are not part of the discriminant variety.  Consider the situation where we have three polynomials $f$, $g$ and $h$ and where $f$ can be factored to $f_1f_2$. One step of the \texttt{ResChainSimple} algorithm will generate two lists. The first one contains $\res(f_1,g,x_N)$ and $\res(f_2,g,x_N)$, and the second one consists of $\res(f_1,h,x_N)$ and $\res(f_2,h,x_N)$. Then in the second step it creates a list containing four resultants in which we can encounter cases such as
\[
\res\Big(\res(f_1,g,x_N),\res(f_2,h,x_N),x_{N-1}\Big).
\]
This will produce an extra component because this is encoding $V(f_1,g)\cap V(f_2,h)$ but we do not need $f_1$ and $f_2$ to both necessarily vanish as they are factors of the same polynomial of the input system $f=f_1f_2$. We only need $f=0$ which means $f_1=0$ \emph{or} $f_2=0$.  A condition on parameters for when both vanish together is not a mandatory condition to have $f=g=h=0$. 

To avoid computing these extra resultants we modify \Cref{Alg:ResChainSimple} to use a further branching idea.  
This time, for each factor of the first polynomial in a branch, we create a new branch for the next step and inside put resultants of that factor with factors of the other polynomials in the lists of that branch. We implemented this algorithm in Maple and called it \texttt{ResChainBranching}. This algorithm is the same as \Cref{Alg:ResChainSimple} with lines $16-25$ replaced with the lines in \Cref{Alg:ResChainBranching}.  In comparison with \texttt{ResChainSimple}, it avoids computing some unnecessary resultants, at the expense of having more branches to keep track of.

\begin{algorithm}
	\caption{A modified version of \Cref{Alg:ResChainSimple} such that some of the extra resultant computations are avoided by replacing the lines $16-25$ of \Cref{Alg:ResChainSimple} with the lines below.}
	\label{Alg:ResChainBranching}
	\begin{algorithmic}[1]
		\For{$f$ in $P_1$}
		\State Let $B'=(\;)$.
		\For{$j$ \textbf{from} $1$ \textbf{by} $1$ \textbf{to} $m-1$}
		\State Let $P_j'=(\;)$.
		\For{$g$ in $P_{j+1}$}
		\If{$f=g$}
		\State Let $B''$ be a list resulted by removing all $P_\ell$ containing $f$ and adding the list $(f)$ at its beginning.
		\State Append $B''$ to $L$.
		\Else
		\State Append all irreducible factors of $\res(f,g,x_i)$ to $P_j'$.
		\EndIf
		\EndFor
		\State Append $P_j'$ to $B'$.
		\EndFor
		\State Append $B'$ to $L'$.
		\EndFor
	\end{algorithmic}
\end{algorithm}

\subsection{Example comparing \texttt{ResChainSimple} and \texttt{ResChainBranching}}

Consider the following simple system of parametric equations of three polynomials, two variables, $x$ and $y$, and a single parameter $a$:
\[
(x^2+y^2-1)(y-x^2)\;=\;x^3+x^2-y^2\;=\;y-a.
\]
To find conditions on $a$ for which this system has a solution we must eliminate the two variables.

The \texttt{ResChainSimple} algorithm gives us the 5 polynomials in the set below.
\[\lbrace a, a+1, a-1, a^2-3a+1, a^4+a^2-1\rbrace.\]
However, the \texttt{ResChainBranching} algorithm gives us only three of them, excluding $a+1$ and $a-1$. Using a Gr\"obner basis computation we get a single polynomial which has three irreducible factors, as from the result of \texttt{ResChainBranching}.  So in this example \texttt{ResChainBranching} did not produce a spurious component, while \texttt{ResChainSimple} produced two extra components because of the extra resultants. 

\subsection{Potential for further optimisation}
\label{sec:resultant_chain_further}

Although \Cref{Alg:ResChainBranching} can avoid some of the unnecessary components provided by \Cref{Alg:ResChainSimple}, it is not guaranteed to produce a minimal number. 

Consider the situation where we have four polynomials $f$, $g$, $h$ and $p$ where $g$ and $h$ can be factored respectively to $g_0q$ and $h_0q$.
The first step of \texttt{ResChainBranching} generates a single branch with three polynomials:
\begin{align*}
	& \res(f,g_0,x_N)\cdot\res(f,q,x_N),\quad \res(f,h_0,x_N)\cdot\res(f,q,x_N),\quad \res(f,p,x_N).
\end{align*}
The first and the second polynomials have a common factor, $\res(f,q,x_N)$, so the algorithm creates a new branch when encountering the request for resultant of this repeated factor from the two polynomials. However, it still computes two unnecessary resultants in the old branch shown below.
\begin{align*}
	& \res\Big(\res(f,g_0,x_N),\res(f,q,x_N),x_{N-1}\Big), \\
	& \res\Big(\res(f,q,x_N),\res(f,h_0,x_N),x_{N-1}\Big).
\end{align*}
It could instead simplify the initial inputs to avoid computing these two extra resultants using the following analysis:
\begin{align*}
	\pi_{N-2}^{N-1}\big(V(f,g,h,p)\big) &= \pi_{N-2}^{N-1}\big(V(f,g_0,h_0,p)\cup V(f,q,p)\big)\\
	&= \pi_{N-2}^{N-1}\big(V(f,g_0,h_0,p)\big)\bigcup\pi_{N-2}^{N-1}\big(V(f,q,p)\big).
\end{align*}
We will leave the quest for finding the most optimised version of \texttt{ResChain} algorithms for future work. We finish this section by presenting an example where the above scenario actually happens.

\subsubsection*{Example} Consider the following simple system of parametric equations of four polynomials, two variables, $x$ and $y$, and three parameters $a$, $b$ and $c$. To find conditions on the parameters for which this system has a solution we must eliminate the two variables.
\[
x^2+y^2-1 \; = \; (x^2-y^2)(x-c) \; = \; (y-x^2+a)(x-c) \; = \; y-b.
\]
The \texttt{ResChainBranching} algorithm returns five polynomials:
\[
\Big\{
c^4-2ac^2+a^2+c^2-1,
c^2-\tfrac{1}{2},
a^2-a-\tfrac{1}{4},
b^2-\tfrac{1}{2},
b^2+c^2-1
\Big\}.
\]
Using the alternative simplified input, i.e. considering first $\lbrace x^2+y^2-1, x^2-y^2, y-x^2+a, y-b\rbrace$ and then considering $\lbrace x^2+y^2-1, x-c, y-b\rbrace$, and then taking the union of the outputs, we get only the last three of the above polynomials. The two polynomials that we do not get are exactly the two resultants mentioned above. To check the validity of the answer, i.e. that the projection is still a subset of the union of solution set of only the three polynomials in the alternative approach, we used elimination via Gr\"obner basis computation. The result contains the following two polynomials.
\[
\Bigg\lbrace 2\Big(b^2-\frac{1}{2}\Big)(b^2+c^2-1), 2\Big(a+b-\frac{1}{2}\Big)(b^2+c^2-1) \Bigg\rbrace.
\]
The variety of the set of the above two polynomials is the union of $V(b^2+c^2-1)$ and two lines defined by $b^2-\frac{1}{2}=a+b-\frac{1}{2}=0$ which are included in $V(a^2-a-\frac{1}{4})\cup V(b^2-\frac{1}{2})$. This shows indeed two polynomials of the result of \texttt{ResChainBranching} are unnecessary output for this example.

\section{Using res-chain with triangular sets}
\label{ResChain_and_triangular_sets}

This is not the first time using a chain of univariate resultants has been suggested. In fact the triangular sets community have noticed its potential earlier. In this section we will first recap the prior work by this community on border polynomials, before using this to propose an improved res-chain approach to our problems.

\subsection{Prior work with border polynomials}

We start with the key definitions.

\begin{definition}
	\label{def:triangular_set}
	Let $X=(x_1,\dots,x_n)$ be a set of ordered indeterminates called variables and $K=\lbrace k_1,\dots,k_r\rbrace$ a set of indeterminates not necessarily ordered, called parameters. For a polynomial $f\in\R[K][X]$, the \emph{leading variable} of $f$ is the variable appearing in $f$ with the highest importance in $X$. Therefore, here the leading variable of $f$ is the $x_i$ with the greatest index appearing in $f$, which we denote by $\lv(f)$. An ordered set of polynomials $F=(f_1,\dots,f_s)$, $s\leq n$ is called \emph{triangular} with respect to $X$ if for every $i<j$ we have $\lv(f_i)<\lv(f_j)$. We call $F$ a \emph{full triangular set} if $s=n$. Trivially a zero dimensional triangular set, meaning that for every specification of the parameters $V(F)$ is a finite set of points, must be a full triangular set.
\end{definition}

\begin{definition}
	\label{def:triangular_decomposition}
	Let $X=(x_1,\dots,x_n)$ and $K=\lbrace k_1,\dots,k_r\rbrace$ be a set of ordered variables and a set of parameters and $F$ a set of polynomials in $\R[K][X]$. An ordered set of triangular sets $(T_1,\dots,T_t)$ together with an ordered set of polynomial sets from $\R[K]$, $(G_1,\dots,G_t)$ is called a \emph{triangular decomposition} for $F$ if for every parameter value $k^\star\in\R^r$ we have $V(F\mid_{k=k^\star})=\cup\big(V(T_i\mid_{k=k^\star})\mid 1\leq i\leq t\text{ such that }k^\star\not\in\cup_{g\in G_i}V(g)\big)$. We call each $T_i$ a \emph{triangular branch} of $F$. Trivially, for every specification of the parameters, by dropping the inequation conditions on parameters, we have $V(F)\subseteq\cup_{i=1}^t V(T_i)$.
\end{definition}

There are different ways to compute a triangular decomposition such as using a Gr\"obner basis or resultant methods, Maple already has an implementation with the command \texttt{Triangularize} in \texttt{RegularChains} package \cite{MorenoMaza-2000}.  We now define the idea of a resultant being taken between a polynomial and a triangular set.

\begin{definition}
	\label{def:resultant_with_respect_to_a_triangular_set}
	\cite[Definition 1.2]{Yang-Hou-Xia-2001} Let $X=(x_1,\dots,x_n)$ and $K=\lbrace k_1,\dots,k_r\rbrace$. Let $g\in\R[K][X]$ and $T=(f_1,\dots,f_n)\subset\R[K][X]$ be a full triangular set. Define $g_n=g$ and then recursively define $g_i=\res(g_{i+1},f_{i+1},x_{i+1})$ for $i=0,\dots,n-1$. Then $g_0$ belongs to $\R[K]$: we call it the resultant of $g$ with respect to $T$ and denote it by $\res(g,T,X)$.
\end{definition}

We now define the \rm{border polynomial}, whose variety contains requested component(s) of the minimal discriminant variety, and therefore is a discriminant polynomial (see Section~\ref{sec:Restricting_to_positive_solutions}). Due to the different approach of computation, it is usually considered as a competitor of the discriminant polynomial calculated following the algorithm of \texttt{RootFinding:-Parametric}. This concept is introduced in \cite[Definition 1.4]{Yang-Hou-Xia-2001}.

\begin{definition}
	\label{def:border_polynomial}
	\cite[Definition 1.4]{Yang-Hou-Xia-2001} Let $X=(x_1,\dots,x_n)$ and $K=\lbrace k_1,\dots,k_r\rbrace$ be variables and parameters and $T=(f_1,\dots,f_n)\subset\R[K][X]$ a full triangular set. The \emph{border polynomial} for the system of equations $f_1=\dots=f_n=0$ is denoted by $\bp(T)$ and is defined as the following product:
	\begin{multline*}
		\res \left( f_1,\frac{df_1}{dx_1},x_1 \right) \cdot 
		\res \left( \res \left(f_2,\frac{df_2}{dx_2},x_2\right), 
		( f_1 ), ( x_1 ) \right) \cdots \\
		\cdot \res \left( \res\left( f_n,\frac{df_n}{dx_n},x_n\right),( f_1,\dots,f_{n-1} ), ( x_1,\dots,x_{n-1} ) \right)
	\end{multline*}
	Trivially $\bp(T)$ belongs to $\R[K]$.
\end{definition}

Theorem 4.2 in \cite{Yang-Hou-Xia-2001} proves that if $\bp(T)\neq 0$, then the solution set of the border polynomial contains the boundary between different parameter regions where the number of real solutions of the system $T=0$ changes; thus it may play the same role as the discriminant variety. 
If one is interested in the change of the number of solutions satisfying an inequality, such as positivity of coordinates or in general $g>0$ where $g\in\R[K][X]$, then the border polynomial needs the extra factor $\res(g,T,X)$ \cite[Definition 1.4, Theorem 4.2]{Yang-Hou-Xia-2001}. Computation of border polynomials is already implemented in Maple by the command \texttt{BorderPolynomial} from the \texttt{ParametricSystemTools} subpackage of \texttt{RegularChains} package.

\subsection{New approach combining triangular sets and chain of resultants}

Forgetting about inequalities, just considering the system $T=0$ for a full triangular system, the border polynomial needs $\sum_{i=1}^ni=\frac{n(n+1)}{2}$ univariate resultants to be calculated, where each of those is obtained by computing a determinant. Alternatively we may use the res-chain idea with respect to a triangular set, but still stick to our initial  approaches using the determinant of Jacobian. Because of the triangular property, the chain of resultants that we introduced in Subsections \ref{sec:Using_resultant_chain}$-$\ref{sec:resultant_chain_further} can be simplified and instead of $i$ resultants in the step where we are eliminating the variable $x_i$, we can simply have just 1 resultant. Note that if we denote the determinant of the Jacobian matrix of $T$ with $d$, then $\pi_{r}^{n+r}\big(V(d)\cap V(T)\big)$, i.e. the projection of the singular points of $V(T)$ on the parameter space, is a subset of $V(\res(d,T,X))$. This will only contain $n$ resultants, and including the calculation of $d$ itself, we use $n+1$ determinants which is less than the number of determinants needed in either of the previous chain of resultants or for the border polynomial. Hence, we introduce a third algorithm naming it \texttt{ResChainTriangular}, see Algorithms~\ref{Alg:ResChainTriangular}. 

In line 3 we use Maple's \texttt{Triangularize} command which takes two inputs, a set of polynomials $F$ and a polynomial ring $R$ which includes the variables as an ordered set $X$ to calculate branches of the triangular decomposition of $F$ with respect to it. Although the \texttt{RegularChains} package has the option of defining a parametric polynomial ring\footnote{\url{https://www.maplesoft.com/support/help/Maple/view.aspx?path=RegularChains/PolynomialRing}}, the \texttt{Triangularize} command will not treat parameters of the input ring in the way we consider parameters; instead it treats them as transcendental numbers meaning that they can not satisfy any polynomial equations\footnote{\url{https://www.maplesoft.com/support/help/maple/view.aspx?path=RegularChains\%2FTriangularize}}. Therefore to use this command in our implementation we have to use polynomial ring $R=\mathbb{R}[K,X]$ with variable ordering of $k_1<\dots<k_r<x_1<\dots<x_n$, i.e. variables being greater than parameters. 

\begin{algorithm}[ht!]
	\caption{Combining the res-chain approach with triangularisation of a set of equations, $F=0$, to eliminate variables from the parametric system of equations $g=F=0$, where $g$ is a polynomial outside of $F$. It can be considered as a generalisation of resultant with respect to full triangular sets.}
	\label{Alg:ResChainTriangular}
	\begin{algorithmic}[1]
		\Procedure{\texttt{ResChainTriangular}}{$g$, $F$, $X=(x_1,\dots,x_n)$, $K=\lbrace k_1,\dots,k_r\rbrace$}
		\Inputo $g$ is a polynomials, $F$ is a list of polynomials, $X$ is an ordered set of $n$ variables (indeterminates to be eliminated) and $K$ is a set of $r$ parameters (the rest of indeterminates).
		\Outputo $S$, a set of irreducible polynomials none of which involve the variables in $X$. The projection of $V(g)\cap V(F)$ as a subset of $\mathbb{R}^{r+n}$ into $\mathbb{R}^r$, is a subset of $\cup_{p\in S}V(p)$.
		\State \emph{Initialisation:} Let $S=\lbrace\;\rbrace$.
		\State Calculate the triangularisation of $F=0$ in the ring $\mathbb{R}[K,X]$ with $k_1<\dots<k_r<x_1<\dots<x_n$ using Maple's \texttt{Triangularize}, say $\big( ( T_1,\dots,T_t ), ( G_1,\dots,G_t ) \big)$.
		\For{$i$ from 1 by 1 to $t$}
		\State Let $T'=\lbrace f\in T_i\mid\lv(f)\in X\rbrace$.
		\State Let $X'=\lbrace\lv(f)\mid f\in T'\rbrace$.
		\If{$T\setminus T'\neq\emptyset$}
		\State Update $S$ with its union with the set of irreducible factors of a polynomial from $T\setminus T'$.
		\Else 
		\State Let $h=\res(g,T',X')$.
		\If{$h\not\in\mathbb{R}[K]$}
		\State Return $\lbrace 0\rbrace$.
		\Else
		\State Update $S$ with its union with the set of irreducible factors of $h$.
		\EndIf
		\EndIf
		\EndFor    
		\State \Return $S$.
		\EndProcedure
	\end{algorithmic}
\end{algorithm}\smallskip

\begin{thm}\label{Thm:correctness_of_algorithm_generalised_resultant}
	Let $X$ and $K$ be an ordered set of $n$ variables and a set of $r$ parameters, $g\in\mathbb{Q}[K][X]$ a polynomial and $F\subset\mathbb{Q}[K][X]$ a set of polynomials. If $S\subset\mathbb{Q}[K]$ is the output of Algorithms~\ref{Alg:ResChainTriangular}, then we have
	\begin{equation}
		\label{Eq:generalised_resultant_with_respect_to_a_set}
		\pi_r^{n+r}\big(V(g)\cap V(F)\big)\subseteq \bigcup_{f\in S}V(f).
	\end{equation}
\end{thm}

\begin{proof}
	Let $\big((T_1,\dots,T_t),(G_1,\dots,G_t)\big)$ be the triangular decomposition of $F$ with respect to $(k_1<\dots<k_r<x_1<\dots<x_n)$ obtained by Maple's \texttt{Triangularize}. For each triangle $T_i$, define $X_i=\lbrace \lv(f)\mid f\in T_i\rbrace\cap X$, $T_i'=\lbrace f\mid f\in T_i, \lv(f)\in X\rbrace$ and $T_i''=T_i\setminus T_i'$. Let $h_i$ to be $\res(g,T_i',X_i)$. If $X_i=X$, we have $h_i\in\mathbb{R}[K]$. But simply because $X_i\neq X$, we can not conclude that $h_i$ would contain a variable, for example it might be the case that $g$ does not involve the variables in $X\setminus X_i$. 
	
	Consider a branch where $h_i\not\in\mathbb{R}[K]$ and $T_i''=\emptyset$. Let $m_i$ be the number of elements in $X_i$, or the number of elements in $T_i'$ and $T_i$. We have
	\begin{align*}
		\pi_r^{n+r}\big(V(g)\cap V(T_i)\big) &= \pi_r^{(n-m_i)+r}\Big(\pi_{(n-m_i)+r}^{n+r}\big(V(g)\cap V(T_i)\big)\Big)\\
		&\subseteq \pi_r^{(n-m_i)+r}\big(V(h_i)\big).
	\end{align*}
	Clearly the projection $\pi_r^{(n-m_i)+r}\big(V(h_i)\big)$ would be $V(0)$ over complex space. Thus if any of such branches appear, Algorithm~\ref{Alg:ResChainTriangular} returns $S=\lbrace 0\rbrace$. In this case relation \eqref{Eq:generalised_resultant_with_respect_to_a_set} trivially holds. For the rest of the proof we assume for any of the branches either $h_i\in\mathbb{R}[K]$ or $T_i''\neq\emptyset$.
	
	Consider a branch where $h_i\not\in\mathbb{R}[K]$ and $T_i''\neq\emptyset$. Let $m_i$ be the number of elements in $X_i$, or the number of elements in $T_i'$. We have
	\begin{align*}
		\pi_r^{n+r}\big(V(g)\cap V(T_i)\big) &= \pi_r^{(n-m_i)+r}\Big(\pi_{(n-m_i)+r}^{n+r}\big(V(g)\cap V(T_i')\big)\cap V(T_i'')\Big)\\
		&\subseteq \pi_r^{(n-m_i)+r}\big(V(h_i)\cap V(T_i'')\big)\\
		&\subseteq V(T_i'').
	\end{align*}
	Define $S_i=T_i''$ and note that $V(T_i'')=\bigcap_{f\in T_i''}V(f)\subseteq\bigcup_{f\in T_i''}V(f)$. So for this branch we proved that
	\begin{equation}
		\label{Eq:Generalised_res_proof_eq_1}
		\pi_r^{n+r}\big(V(g)\cap V(T_i)\big) \subseteq \bigcup_{f\in S_i}V(f).
	\end{equation}
	
	Consider a branch where $h_i\in\mathbb{R}[K]$ and $T_i''=\emptyset$. We have
	\[\pi_r^{n+r}\big(V(g)\cap V(T_i)\big) \subseteq V(h_i).\]
	Then by defining $S=\lbrace h_i\rbrace$, relation \eqref{Eq:Generalised_res_proof_eq_1} holds for this branch as well.
	
	Finally consider a branch where $h_i\in\mathbb{R}[K]$ and $T_i''\neq\emptyset$. In this case $X_i=X$ but number of polynomials in $T_i$ is bigger than $n$ (but not bigger than $n+r$). We have
	\begin{align*}
		\pi_r^{n+r}\big(V(g)\cap V(T_i)\big) &= \pi_r^{n+r}\big(V(g)\cap V(T_i')\cap V(T_i'')\big)\\
		&= \pi_r^{n+r}\big(V(g)\cap V(T_i')\big)\cap V(T_i'')\\
		&\subseteq V(h_i)\cap V(T_i'')\\
		&= V(T_i''\cup\lbrace h_i\rbrace).
	\end{align*}
	Then by defining $S=T_i''\cup\lbrace h_i\rbrace$, relation  \eqref{Eq:Generalised_res_proof_eq_1} holds again.
	
	Since relation \eqref{Eq:Generalised_res_proof_eq_1} holds for all branches, we can deduce that
	\begin{align*}
		\pi_r^{n+r}\big(V(g)\cap V(F)\big) &\subseteq \pi_r^{n+r}\Big( V(g)\cap \big(\bigcup_{i=1}^t V(T_i)\big)\Big)\\
		&= \pi_r^{n+r}\Big(\bigcup_{i=1}^t\big(V(g)\cap V(T_i)\big)\Big)\\
		&= \bigcup_{i=1}^t\pi_r^{n+r}\big(V(g)\cap V(T_i)\big)\\
		&\subseteq \bigcup_{i=1}^t\bigcup_{f\in S_i}V(f).
	\end{align*}
	Next, by defining $S=\cup_{i=1}^tS_i$, relation \eqref{Eq:generalised_resultant_with_respect_to_a_set} holds. Note that if for every branch, instead of all selected polynomials, only one of them is picked, the set inclusion still remains valid. Additionally, replacing polynomials in $S$ by their irreducible factors do not change the result.  This completes the proof.    
\end{proof}

\begin{remark}
	\label{Rem:Generalised_res_output_being_SOS}
	Consider the assumptions of Theorem~\ref{Thm:correctness_of_algorithm_generalised_resultant} and all notations introduced in its proof. Define $A$ and $B$ to be subsets of $\lbrace 1,\dots,t\rbrace$ corresponding to the indices of branches where $h_i\not\in\mathbb{R}[K]$ and $T_i''=\emptyset$ and indices of branches where $h_i\not\in\mathbb{R}[K]$ and $T_i''\neq\emptyset$, respectively. Note that for a branch $T_i$ that $i\not\in A\cup B$, we have $h_i\in\mathbb{R}[K]$ and $T_i''$ might be empty or nonempty. The proof of Theorem~\ref{Thm:correctness_of_algorithm_generalised_resultant} actually proves two better results.
	\begin{itemize}
		\item[i)] If $A=\emptyset$, then
		\[\pi_r^{n+r}\big(V(g)\cap V(F)\big) \subseteq \Big(\underset{\substack{1\leq i\leq t\\i\not\in B}}{\bigcup}\big(V(h_i)\cap V(T_i'')\big)\Big)\cup\Big(\bigcup_{i\in B}V(T_i'')\Big).\]
		\item[ii)] For every $i\in B$, define $f_i=\sum_{f\in T_i''}f^2$, and for every $1\leq i\leq t$ that $i\not\in B$, define $f_i=h_i^2+\sum_{f\in T_i''}f^2$. Let $S=\lbrace f_1,\dots,f_t\rbrace$. Over the reals we have that $\bigcup_{f\in S}V(f)$ is equal to $\Big(\underset{\substack{1\leq i\leq t\\i\not\in B}}{\bigcup}\big(V(h_i)\cap V(T_i'')\big)\Big)\cup\Big(\bigcup_{i\in B}V(T_i'')\Big)$, so
		\[\pi_r^{n+r}\big(V(g)\cap V(F)\big) \subseteq \bigcup_{f\in S}V(f).\]
		This is a sharper projection than the one given by Algorithm~\ref{Alg:ResChainTriangular}. However if size of polynomials is a concern, the output of Algorithm~\ref{Alg:ResChainTriangular} would be preferred.
	\end{itemize}
\end{remark}

As a simple example consider $X=(x,y)$, $K=\lbrace a,b\rbrace$, $g=x^2+y^2-1$ and five triangular sets over $\mathbb{R}[a,b,x,y]$ with $a<b<x<y$ as follows
\begin{align*}
	T_1&=(x-a,y-b), \quad T_2=(a-1,x-y), \quad T_3=(a-1,b-1), \\
	T_4&=(b,x-a,x-y), \quad T_5=(x-a).
\end{align*}
The triangular decomposition of $F=\lbrace f_1f_2f_3f_4f_5\mid f_i\in T_i,i=1,\dots,5\rbrace$ calculated by Maple's \texttt{Triangularize} is $\big((T_1,\dots,T_5),(\emptyset,\dots,\emptyset)\big)$. One can do the calculations of Algorithm~\ref{Alg:ResChainTriangular} by hand for this example: $\res(g,T_5,(x))=y^2+a^2-1$ and $T_5\setminus T_5'=\emptyset$, so the algorithm returns the projection to be $\pi_2^4\big(V(g)\cap V(F)\big)\subseteq V(0)$. Using the \texttt{EliminationIdeal} command of Maple to eliminate $x$ and $y$ from the ideal generated by $F\cup\lbrace g\rbrace$, we get $\langle 0\rangle$, and thus the same conclusion. 

Redefining $F$ to be $F=\lbrace f_1f_2f_3f_4\mid f_i\in T_i,i=1,\dots,4\rbrace$, the triangular decomposition becomes  $\big((T_1,\dots,T_4),(\emptyset,\dots,\emptyset)\big)$. For the first branch we have $\res(g,T_1,X)=a^2+b^2-1$ so $\pi_2^4\big(V(g)\cap V(T_1)\big)\subseteq V(a^2+b^2-1)$. For the second branch we have $\res\big(g,(x-y),(y)\big)=2x^2-1$ and $T_2\setminus T_2'=\lbrace a-1\rbrace$, so $\pi_2^4\big(V(g)\cap V(T_2)\big)\subseteq \pi_2^3\big(V(2x^2-1)\cap V(a-1)\big)\subseteq V(a-1)$. For the third branch we have $T\setminus T'=T$, so $\pi_2^4\big(V(g)\cap V(T_3)\big)\subseteq V(T_3)$. For the fourth and the last branch we have $\res\big(g,(x-a,x-y),(x,y)\big)=2a^2-1$ and $T\setminus T'=\lbrace b\rbrace$, so $\pi_2^4\big(V(g)\cap V(T_4)\big)\subseteq V(2a^2-1,b)$. By Remark~\ref{Rem:Generalised_res_output_being_SOS}(i) we have 
\[\pi_2^4\big(V(g)\cap V(F)\big)\subseteq V(a^2+b^2-1)\cup V(a-1)\cup V(a-1,b-1)\cup V(2a^2-1,b).\]
The third component of course is subset of the second one and can be ignored. The output of Algorithm~\ref{Alg:ResChainTriangular} is $\cup\big(V(f)\mid f=a^2+b^2-1,a-1,b\big)$ which is a superset. Using the \texttt{EliminationIdeal} command of Maple we get an ideal generated by the following three polynomials:
\begin{align*}
	&\{ a^4b + a^2b^3 - 2a^3b - 2ab^3 + b^3 + 2ab - b, -2a^5 + 2ab^4 + 2a^4 - a^3b - ab^3 \\
	&\qquad - 2b^4 + 3a^3  + a^2b - ab^2 + b^3 - 3a^2 + ab + b^2 - a - b + 1, 2a^5 + 2a^3b^2 \\
	&\qquad- 2a^4 - a^3b - 2a^2b^2 - ab^3 - 3a^3 + a^2b - ab^2 + b^3 + 3a^2 + ab + b^2 + a - b - 1\}.
\end{align*}
Checking the triangular decomposition of this variety, we get 
\[\
Big(\big((a^2+b^2-1),(a-1),(2a^2-1,b)\big),\big(\emptyset,\emptyset,\emptyset\big)\Big),
\]
which shows that in this case, Remark~\ref{Rem:Generalised_res_output_being_SOS}(i) provides the exact sharp projection. 
One can simply check that removing $T_2$ from definition of $F$ will result in getting $\pi_2^4\big(V(g)\cap V(F)\big)\subseteq V(a^2+b^2-1)\cup V(a-1,b-1)\cup V(2a^2-1,b)$ and the result is the sharp exact projection, i.e. equality occurs.


Note that the border polynomial command in Maple computes the border polynomial for non-triangular sets such as $F$, the input for our \texttt{ResChainTriangular}, but it is not just the product of border polynomials for the triangular branches.  For example, it also takes into account when different branches can have intersections. We do not need to worry about these intersections in \texttt{ResChainTriangular} since the determinant of the Jacobian of $F$ already contains this information as well.

\newpage

\section{Application of New Approaches to Population Model Application}
\label{sec:Application_of_New_Approach_to_Population_Model}

\subsection{The three-dimensional case}

Consider System \eqref{Eq:n-patch-ODE} with $n=3$ and denote the polynomials on the right by $f_i$, $i=1,2,3$. Let $d$ be the determinant of the Jacobian matrix of $f=(f_1,f_2,f_3)$ with respect to $x=(x_1,x_2,x_3)$. Then the ideal associated with the discriminant variety of this system is 
$
\langle f_1,f_2,f_3,d\rangle\cap\mathbb{R}[a,b].  
$  
Recall from \Cref{sec:Recent_Prior_work_on_Population_Model_Application} that we could not before study this symbolically when computing the discriminant variety with a simple use of Gr\"{o}bner basis, and thus relied on a symbolic-numeric analysis instead.

Now, instead of using GB to find a basis for this elimination ideal, we may compute the Dixon resultant of the polynomial set $\lbrace f_1,f_2,f_3,d\rbrace$ with respect to the variables $x$. It took about 6 minutes on our laptop and the result is the polynomial below, which has  8 irreducible factors, two of which have no solutions in the positive orthant (namely $3a+b$ and $2a+b$).
\begin{equation*}\label{eq:Dixon_resultant_three_populations}
	\begin{array}{l}
		42391158275216203514294433201 \cdot b^{2} \cdot (3 a +b )^{8} \cdot (2 a +b )^{24} \cdot (4 a \,b^{4} \\ \quad
		-36 a^{2} b^{2}-8 a \,b^{3}-b^{4}+108 a^{3}+36 a^{2} b+12 a \,b^{2}+2 b^{3}-36 a^{2}-8 a b - b^{2}+4 a )^{9} \\
		\cdot \, (256 a^{4} b^{10}-32 a^{2} b^{12}-6144 a^{5} b^{8}-1280 a^{4} b^{9}+
		768 a^{3} b^{10} + 192 a^{2} b^{11}+12 a \,b^{12}\\ \quad+54784 a^{6} b^{6}+24576 a^{5} b^{7}-
		4416 a^{4} b^{8}- 3840 a^{3} b^{9}-776 a^{2} b^{10}- 72 a \,b^{11}-b^{12}\\ \quad
		- 165888 a^{7} b^{4}-164352 a^{6} b^{5} - 16512 a^{5} b^{6}+25344 a^{4} b^{7}+10848 a^{3} b^{8}+2120 a^{2} b^{9}\\ \quad
		+204 a \,b^{10}+6 b^{11} - 248832 a^{8} b^{2}+331776 a^{7} b^{3}+207744 a^{6} b^{4}-36480 a^{5} b^{5} \\ \quad 
		-54528 a^{4} b^{6}-20352 a^{3} b^{7}-3800 a^{2} b^{8}-360 a \,b^{9} -
		15 b^{10}+2239488 a^{9}+248832 a^{8} b \\ \quad -497664 a^{7} b^{2}-141568 a^{6} b^{3}+
		62976 a^{5} b^{4}+69504 a^{4} b^{5}+25152 a^{3} b^{6} +4592 a^{2} b^{7}\\ \quad +432 a \,b^{8}+20 b^{9}
		-248832 a^{8}+331776 a^{7} b +207744 a^{6} b^{2}-36480 a^{5} b^{3}-54528 a^{4} b^{4} \\ \quad
		- 20352 a^{3} b^{5}-3800 a^{2} b^{6}-360 a \,b^{7}-15 b^{8}-165888 a^{7}-
		164352 a^{6} b -16512 a^{5} b^{2} \\ \quad 
		+25344 a^{4} b^{3}+ 10848 a^{3} b^{4}+2120 a^{2} b^{5}+
		204 a \,b^{6}+6 b^{7}+54784 a^{6}+24576 a^{5} b \\ \quad 
		-4416 a^{4} b^{2}-3840 a^{3} b^{3}-
		776 a^{2} b^{4}- 72 a \,b^{5}-b^{6}-6144 a^{5}-1280 a^{4} b +768 a^{3} b^{2}\\ \quad
		+192 a^{2} b^{3}+12 a \,b^{4}+256 a^{4}-32 a^{2} b^{2})^{3} \\ 
		\cdot \, (3 a +1-b )^{8} \cdot (b^{2}+ 3 a -b )^{8} \cdot (b -1)^{2}
	\end{array}
\end{equation*}

\begin{figure}[tbhp]
	\centering
	\subfloat[]{\label{Fig:ResChainSimple_1}\includegraphics[width=3.9cm]{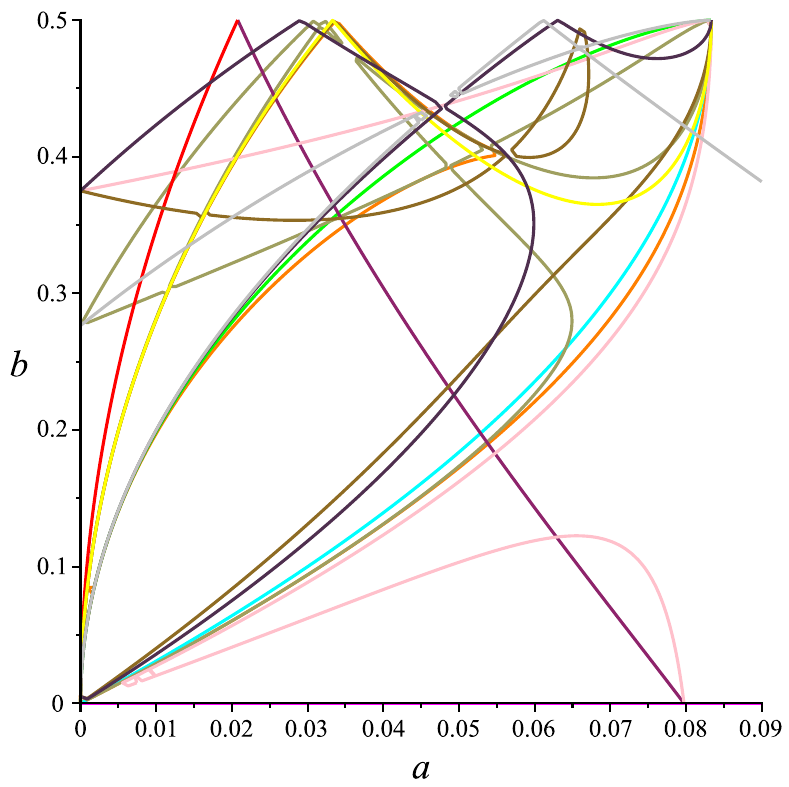}} \qquad
	\subfloat[]{\label{Fig:ResChainSimple_zoomed}\includegraphics[width=3.9cm]{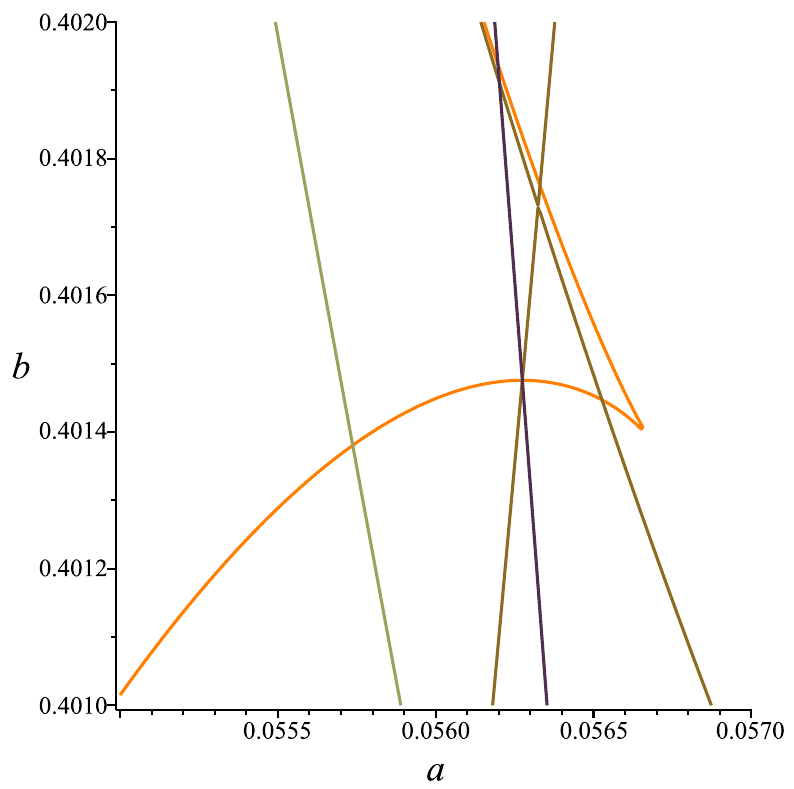}}\\
	\subfloat[]{\label{Fig:ResChainBranching_1}\includegraphics[width=3.9cm]{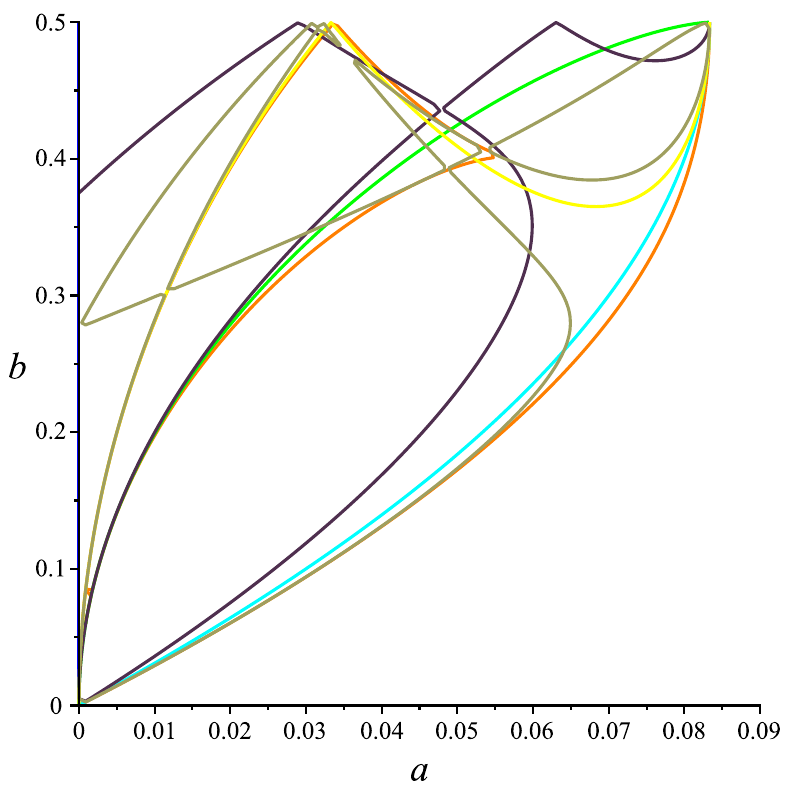}} \qquad
	\subfloat[]{\label{Fig:ResChainBranching_zoomed}\includegraphics[width=3.9cm]{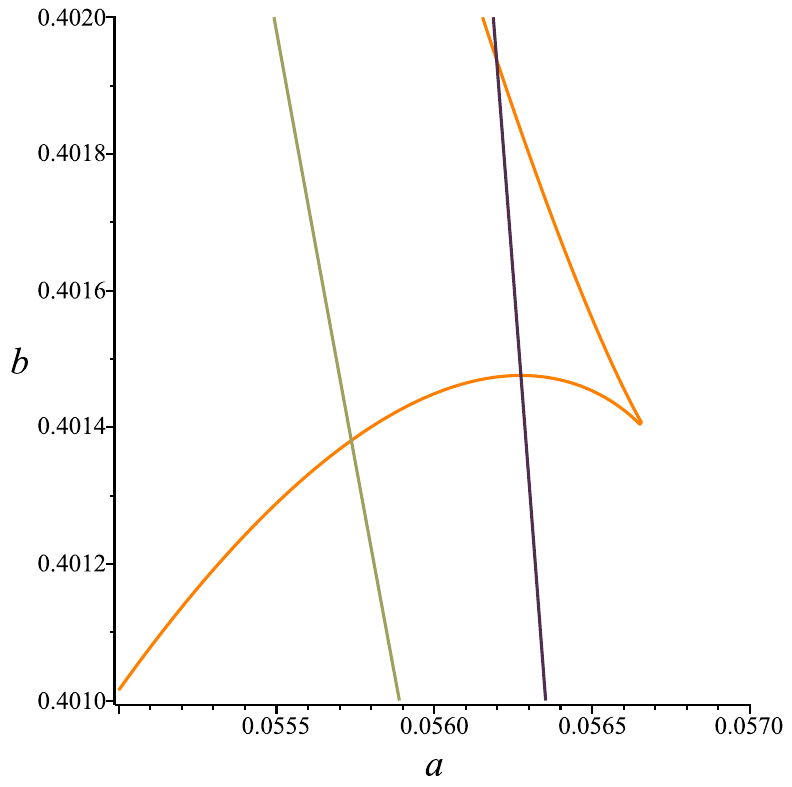}}\\
	\subfloat[]{\label{Fig:Dixon_1}\includegraphics[width=3.9cm]{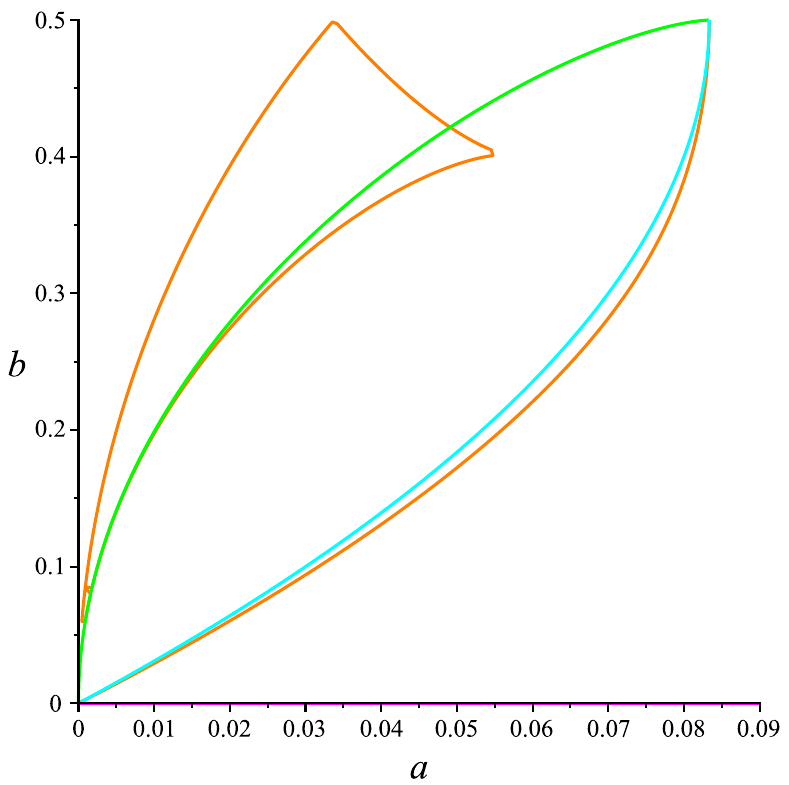}} \qquad
	\subfloat[]{\label{Fig:Dixon_zoomed}\includegraphics[width=3.9cm]{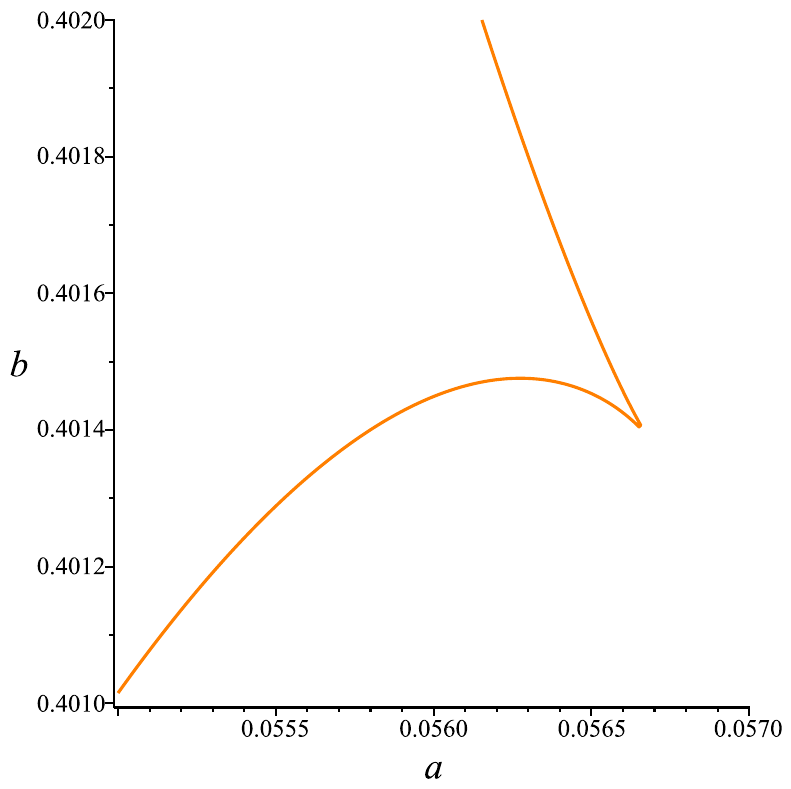}}\\
	\subfloat[]{\label{Fig:BP_1}\includegraphics[width=3.9cm]{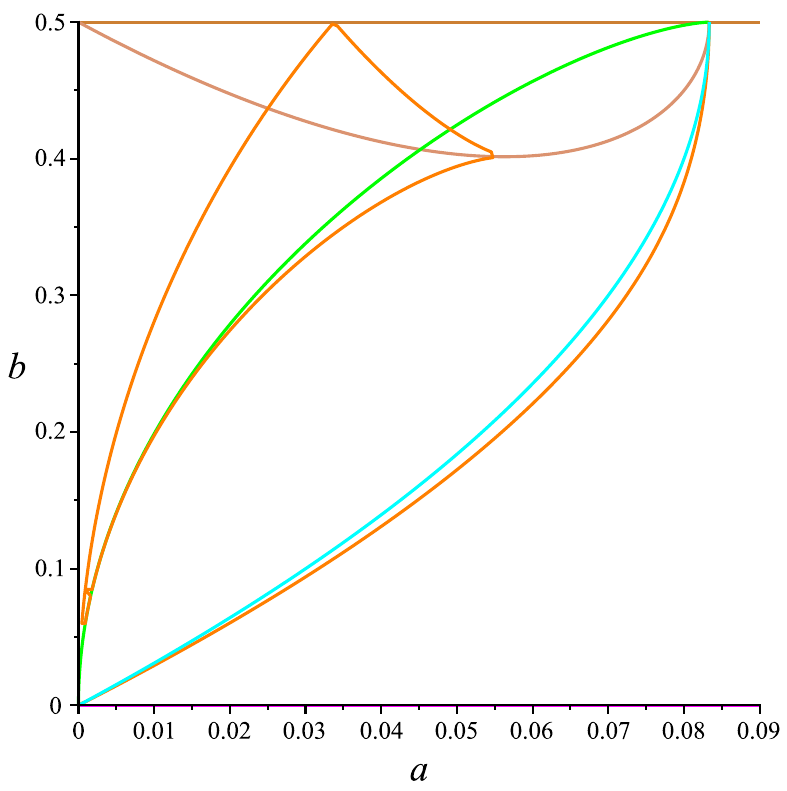}} \qquad
	\subfloat[]{\label{Fig:BP_zoomed}\includegraphics[width=3.9cm]{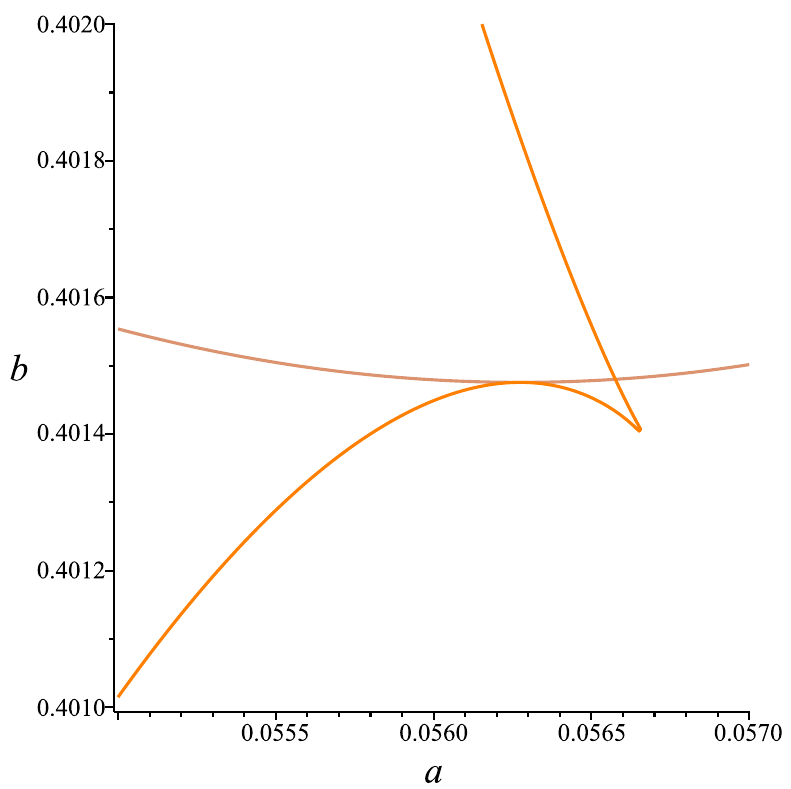}}
	\caption{Graph of the product of the polynomials obtained from eliminating variables $x_1$, $x_2$ and $x_3$ from the system made by $\lbrace f_1,f_2,f_3,d\rbrace$ using four resultant techniques instead of Gr\"obner basis computation (the first three rows) and the border polynomial of the system $f_1,f_2,f_3$ (the last row). The figures in the left column show the region $[0, 0.5]\times [0,\infty)$, and the figures in the right column are zoomed area of their left figures in the region $[0.055, 0.057]\times [0.401, 0.402]$.}
	\label{Fig:new_approach}
\end{figure}

\Cref{Fig:Dixon_1} shows the plot of the solution set of the polynomial found by the Dixon resultant. It is the exact discriminant variety approximated by the symbolic-numeric approach in \Cref{sec:Recent_Prior_work_on_Population_Model_Application}. The interesting region where the number of solutions could temporarily increase is related to the solution set of the largest factor of this polynomial. \Cref{Fig:Dixon_zoomed} shows the zoomed version of this curve at this interesting region. Having only the boundaries that we found by the symbolic-numeric approach in the result of elimination via the Dixon resultant also proves that the result of the Dixon resultant does not contain any extra component in this example, and the result of the symbolic-numeric approach was complete and the behaviour of the system was indeed completely classified in \cite{Rost-Sadeghimanesh-2021-1}.

We also applied our three res-chain algorithms, Algorithms~\ref{Alg:ResChainSimple}$-$\ref{Alg:ResChainTriangular}. They take about half of a second, about 5 milliseconds and about 2 deciseconds respectively. They are much faster than the Dixon resultant, but the first two include additional unnecessary components in the output. The simple version returns 16 irreducible components and the modified one returns 11 components. The set of irreducible factors in the Dixon resultant is a subset of the set of irreducible polynomials in the \texttt{ResChainBranching}, and the latter is a subset of the set of irreducible polynomials in the result of \texttt{ResChainSimple}. \texttt{ResChainTriangular} returns the same 8 components of the Dixon resultant. Figures~\ref{Fig:ResChainSimple_1}$-$\ref{Fig:ResChainBranching_zoomed} show the plots of solution sets of the product of the polynomials in the output of these two \texttt{ResChainSimple} and \texttt{ResChainBranching} methods. Note that the plot for \texttt{ResChainTriangular} would be the same as for the Dixon resultant, i.e. \cref{Fig:Dixon_1} and \cref{Fig:Dixon_zoomed}. 

Finally, the  border polynomial is also calculated, taking about 2 decisecond (the same as \texttt{ResChainTriangular}). The result contains 12 irreducible components, 8 of which are the ones in the result of the Dixon resultant and \texttt{ResChainTriangular}, but the extra 4 ones are new, not even among extra components of \texttt{ResChainSimple} and \texttt{ResChainBranching}. \cref{Fig:BP_1} and \cref{Fig:BP_zoomed} show the plots of the solution sets of the product of the polynomials in the output of border polynomial approach. We coloured each of the 20 components, 16 of \texttt{ResChainSimple} and the extra 4 of border polynomial, with a different colour and we used the same colour for common curves among results of different approaches for a better comparison. The largest polynomial in the output of both \texttt{ResChainSimple} and \texttt{ResChainBranching} approaches has 153 terms and total degree of 21. The largest factor of the result of the the other three approaches has 72 terms and total degree of 14.

The new approaches can all produce information on the discriminant variety which was infeasible when using GB. There is a trade-off between the speed of computation and the presence of redundant components in the output. Depending on the relative sizes of the variable and parameter spaces one approach may be preferred over the other. We note that an open CAD with respect to the union of the polynomials in the output of \texttt{ResChainBranching} (including the unnecessary ones) finishes in about 2 minutes. 
However, the CAD computation for \texttt{ResChainSimple} output did not terminate after an hour. 

\subsection{Higher-dimensional case}

So we have tackled the previously intractable 3-population case; and a natural question is whether these approaches are sufficient for the 4-population case?  The Dixon resultant computation for four connected populations did not terminate after four hours on our laptop, while the first two \texttt{ResChain} algorithms both encounter a branch with a single polynomial and so return $0$. Border polynomials goes up to $n=6$ in a reasonable amount of time but for $n=7$ it throws the following error message about size of the polynomial in Maple.\smallskip 

\begin{verbatim}
	Error, (in Tools:-Regroup) too many levels of recursion
\end{verbatim}\smallskip

We note that we reported on the use of border polynomials in the Maple conference 2022 as part of the art gallery competition, see \cite{Allee-flowers}. \texttt{ResChainTriangular} takes less than 2 seconds for the case $n=4$ and returns less redundant factors than the output of border polynomial, 11 versus 21. Thus we have successfully pushed forward the limit of computation for this problem.

\section{Other Attempts to Tackle the $n$-patches Allee Effect Model in the literature}
\label{sec:Other_groups_attempts}

\subsection{The work of Song and Tang (2026)}

In \cite{Song-Tang-2026}, the authors attempted calculating the discriminant of System \eqref{Eq:n-patch-ODE} using border polynomials.  They observed in \cite[Theorem 3.1]{Song-Tang-2026} that the coordinates of the steady states of this system can involve at most three distinct positive values. Thus, by adding extra equations of the form $x_i=x_j$ they simplify the system to a smaller system of 3 equations in 3 variables. Then they report on computation time of border polynomials after this simplification for $n$ up to 100 versus computation time for border polynomial before this simplification for $n$ up to 6 \cite[Table 1]{Song-Tang-2026}. They failed to acknowledge that the use of border polynomials had already been reported in \cite{Allee-flowers} (three years before the preprint of \cite{Song-Tang-2026} was placed on the arXiv); and in fact the error in Maple for $n=7$ has been reported there as well.

One should note that while \cite{Song-Tang-2026} has focused on a specific problem in a full symmetry, not only assuming complete directed graph of connectivity among different patches, but also having all dispersal rates of all edges being equal.  In contrast, the work in the present paper is not concerned with just one case study and our quest is for better elimination algorithms that work in general. For example in \cite{Rost-Sadeghimanesh-2021-2} a simple example of non-fully symmetric has been considered (unidirectional migration among two patches). There is no reason to expect these symmetries to hold in generic examples from population dynamics or CRN theory.

Let us demonstrate that in the $n$-patches population dynamics with Allee effect, the simplification used in \cite{Song-Tang-2026} does not hold when we move away from the fully symmetric case.  Consider the case $n=4$ where the dispersal rates between different patches is allowed to be different, meaning that the model is not fully symmetric.  Then the system can obtain a solution with all four coordinates different. For example by letting 
\begin{align*}
	a_{12}=a_{13}=a_{14}=0.01, &\quad a_{21}=a_{23}=a_{24}=0.02, \quad a_{31}=a_{32}=a_{34}=0.03,\\
	a_{41}=a_{42}=a_{43}=0.05, &\quad b=0.45,
\end{align*}
where $a_{ij}$ is the dispersal rate from the $i$th patch to the $j$th patch, there are 22 solutions with all 4 coordinates different such as 
\[x=(0.3641901, 0.4592824, 0.6188128, 0.0609088)\]
with seven digits after decimal point of accuracy.

\subsection{The work of Tsai (2026)}

Another paper, \cite{Tsai-2026} attempted the same $n$-patches Allee effect problem, consisting of two parts. Part 1 concerns the full symmetric case with the same observation as in \cite{Song-Tang-2026}, see \cite[Lemma 2]{Tsai-2026}. Part 2 uses our res-chain idea which existed in the initial draft of this manuscript \cite[Algorithms 1]{Resultant-paper-arXiv} in 2022, which they failed to acknowledge. They use a proposition regarding the component of discriminant variety that comes from the singular locus, see \cite[Proposition 3]{Tsai-2026} which states that under a technical condition, projection of a common solution of two polynomials with intersection multiplicity higher than one is a singular solution of resultant of the two polynomials. Therefore they suggest using the res-chain idea on $F=\lbrace f_1,\dots,f_n\rbrace$ with respect to $x=(x_2,\dots,x_n)$, instead of using the res-chain idea on $F=\lbrace f_1,\dots,f_n,d\rbrace$ with respect to $x=(x_1,\dots,x_n)$, calling the output $h$, then calculating $\res(h,\frac{dh}{dx_1},x_1)$. In fact, \cite{Tsai-2026} uses our modified definition of univariate resultant in section~\ref{sec:Using_resultant_chain}, and even points out that if two resultants in the same step of \texttt{ResChainSimple} contain common factors the approach is infeasible because of obtaining a zero polynomial, which is the \texttt{return([0])} line of Algorithm~\ref{Alg:ResChainSimple}.  These were both already present in the first draft of this manuscript available on the arXiv \cite{Resultant-paper-arXiv}. We had discussed the later point in more detail with an attempt to improve the algorithm. Thus the new algorithm in Part 2 of \cite{Tsai-2026} is actually using our res-chain idea on a different input set.

Their work concerns only with the component of discriminant variety coming from singular locus. They move away from the fully symmetry case in the connectivity graph of the $n$-patch population model, in contrast to the work in \cite{Song-Tang-2026}, and motivated from the work in \cite{Rost-Sadeghimanesh-2021-2} where the authors had studied all non-isomorphic connectivity graphs for $n=2$, but this time considering for $n=3$.

However, \cite[Proposition 3]{Tsai-2026} can not be used when dealing with other components of the discriminant variety, for example when restricting solutions to those satisfying a polynomial inequality $g>0$. That means the quest for better elimination algorithms still remains open. So in summary the drawbacks are: limited to only the singular locus component of discriminant; needs checking the conditions for the possibility of ignoring $d$ as an input to a res-chain approach, which \cite{Tsai-2026} suggests use of Gr\"obner basis for each elimination level (thus $n-1$ round of checks); and the problem of hitting \texttt{return([0])} line because of common factors.

\subsection{The work of Breiding et al., (2026)}

Finally, \cite{Breiding-et-al-2026} discusses getting information about the complement of the discriminant variety of a parametric system without calculating a discriminant variety itself. In the present paper we did not compare performance of our algorithm with that approach for two reasons: firstly because it is ignoring the elimination problem and thus not finding the discriminant polynomial itself, and secondly because their approach falls in the category of numerical approaches, using numerical homotopy methods.


\section{Benchmarking on a Wider Set of Examples}
\label{sec:Benchmarking}

The algorithms developed here can be used for any other square zero-dimensional parametric system including a wide range of applications in population dynamics and chemical reaction network theory. In this paper six methods for elimination tasks with $n+1$ polynomials in $n$ variables as the input have been discussed, all with implementations in Maple:
\begin{enumerate}
	\item using Gr\"obner basis \\(\texttt{PolynomialIdeals:-EliminationIdeal}),
	\item using the Dixon resultant \\(\texttt{DR:-DixonResultant}),
	\item using simple chain of resultants \\(\texttt{ResChain:-ResChainSimple}),
	\item using a modification of the simple chain of resultants \\(\texttt{ResChain:-ResChainBranching}),
	\item using Border Polynomials \\(\texttt{RegularChains:-ParametricSystemTools:-BorderPolynomial}),
	\item and using a chain of resultants on the triangular decomposition of the input \\(\texttt{ResChain:-ResChainTriangular}).
\end{enumerate}
To compare performance of all these six approaches we picked 49 systems  and run all six methods to calculate only the part of the discriminant variety that is obtained from projecting the singular locus of the variety, therefore restricting the problem to only calculating one single elimination. The first one, indexed 0, is the simple univariate quadratic equation discussed in Section~\ref{sec:Solution_via_CAD_and_Discriminant_Variety}. The next 35 systems are from CRN theory, in order, 2-site phosphorylation-dephosphorylation cycle \cite{Wang-Sontag-2008}, hybrid histidine kinase \cite{Kothamachu-et-al-2015}, an autoregulatory network \cite[Figure 3B]{Siegal-et-al-2009}, and the 32 fully open CRNs in \cite[Appendix B]{MLCRN}. Finally 13 systems come from population dynamics: the first 6 systems are the fully symmetric $n$-patches population with Allee effect (see Equation~\eqref{Eq:n-patch-ODE}) for $n=2,\dots,7$, similarly the last 6 equations are fully symmetric $n$-patches population with logistic growth for $n=2,\dots,7$. The logistic systems are the same as the Allee effect ones just missing the term $(x_i-b)$. The system indexed 42 is the symmetric 2-patch population with Allee effect and Holling response function type 2 (for a simple definition of Holling response function type 2 see \cite[page 12]{Best-2023}). This system has two variables $x_1$ and $x_2$, populations at patches 1 and 2, and four parameters $a$ the dispersal rate, $b$ the Allee effect threshold, $\rho$ and $A$ the parameters of the Holling response term. The equations are
\[
\left\lbrace\begin{array}{l}
	\dot{x}_1 = x_1(1-x_1)(x_1-b)-\frac{\rho x_1}{x_1+A}-ax_1+ax_2\\
	\dot{x}_2 = x_2(1-x_2)(x_2-b)-\frac{\rho x_2}{x_2+A}-ax_2+ax_1.
\end{array}\right.
\]
Substituting $\dot{x_i}=0$, $i=1,2$ and multiplying both sides of the two equations with $x_i+A$ (note that $A>0$ and $x_i\geq 0$ thus $x_i+A>0$, particularly non-zero) the system becomes of polynomial form suitable to feed to the six algorithms.

The size of the systems varies with $n$, the number of variables, between 1 and 9, $r$, number of parameters, between 1 and 18, $D$, maximum total degree of the polynomials, between 2 and 5. The exact statistics of the size of systems can be found in columns 2 to 4 of Table~\ref{tab:benchmark}. 
We wrote Python code to automatically call a fresh Maple session with no previous data stored in the session from the calculations of other calls and run one of the six algorithms on one of the systems. Each time we measured three different timings: 
\begin{enumerate}
	\item The time measured inside Maple for the elimination algorithm including any preprocessing needed such as calculation of the determinant of the Jacobian matrix.
	\item The time measured inside Maple for both the previously mentioned parts and factorising the polynomial(s) in the result in a predefined format and order so the outputs of all six algorithms can be compared. Some of these algorithms are structured in a way that the output is already factorised so there are factorisation costs paid already.
	\item Finally the time measured by Python from the moment of opening the Maple session until closing the Maple session.
\end{enumerate}
We put a timeout limit of 600 seconds measured in Python. If the calculation is not done before passing this time limit, Python will terminate the Maple session. In Table~\ref{tab:benchmark} we only show the second Maple time. 
Columns 5 to 10 contain this Maple time corresponding to the six methods using Gr\"obner basis (GB), Dixon resultant (DR), \texttt{ResChainSimple} (RS), \texttt{ResChainBranching} (RB), border polynomial (BP) and \texttt{ResChainTriangular} (RT) respectively.
We also recorded the number of irreducible factors of the output, the maximum total degree and the maximum number of terms of these factors (not shown in Table~\ref{tab:benchmark}). These are useful to see which method produced less unnecessary factors and of smaller size. 

Other than timeout, denoted by \texttt{TO} in Table~\ref{tab:benchmark}, there are two error cases that may rise. Method 1, using Gr\"obner basis, triggered the following error message on systems with ID numbers 32, 34 and 35.

\begin{verbatim}
	maple: fatal error, lost connection to kernel
\end{verbatim}

This error message means that Maple was consuming too much resources from our device, most likely RAM\footnote{\url{https://faq.maplesoft.com/s/article/Kernel-connection-has-been-lost-error-in-Maple?language=en\_US}}. This is denoted by \texttt{ER1} in Table~\ref{tab:benchmark}. The second error message occurs for Method 5 when border polynomial has problem with the size of polynomials.  This is the error message mentioned earlier in Section~\ref{sec:Application_of_New_Approach_to_Population_Model}. This is denoted by \texttt{ER2} and happened for systems with ID numbers 41, 46-48 which are the 7-patches Allee effect and $n$-patches logistic growth for $n\geq 5$. \texttt{GU} stands for `give up' and occurs for Methods 3 and 4 when they hit the inconclusive return $\lbrace 0\rbrace$ line. In each row of Table~\ref{tab:benchmark} the timeouts, errors and give up cases are coloured gray and the cell related to the method that took the least amount of time among those that managed to finish the calculation successfully is coloured yellow.

Since bringing the rest of information on the size of outputs of these 6 methods will add extra tables, to save space, we refer the interested reader to the repository of the paper, a link to which can be found in the Data Access Statement of the paper. 
The output of \texttt{ResChainTriangular} most of the time has fewer  factors than the output of border polynomials. Out of 49 systems, for 7 of them neither algorithm was successful in the time limit of 600 seconds. Gr\"obner basis has the highest win in computation time with 17 out of 42 remaining systems, the second method is Dixon resultant with 9 wins, border polynomials and \texttt{ResChainTriangular} both have 6 wins and finally the other two other res-chain routes with 2 wins each. \texttt{ResChainTriangular} usually takes less or the same amount of time compared to border polynomials. One may decide to design a queue of  methods to attempt, i.e. with a second or two timelimit on GB, and then if it failed to finish the computation we attempt the Dixon resultant, and if that failed to complete in the given time limit switch to \texttt{ResChainTriangular}, or a similar idea.

The reason we do not compare these 6 methods with the work in \cite{Tsai-2026} is that the use of \cite[Proposition 3]{Tsai-2026} can be combined with any of Methods $1-4$ and 6. All they need is answering \textbf{Problem 2} for an input with one polynomial and one variable less and in the end taking a univariate discriminant. The goal of Table~\ref{tab:benchmark} is benchmarking the elimination algorithms themselves.

\begin{table}[ht!]
	\centering
	\resizebox*{!}{0.9\textheight}{%
		\begin{tabular}{|c|c|c|c|c|c|c|c|c|c|}
			\hline
			\textbf{idx} &  $\bm{n}$ & $\bm{r}$ & $\bm{D}$ & \textbf{GB} & \textbf{DR} & \textbf{RS} & \textbf{RB} & \textbf{BP} & \textbf{RT}\\
			\hline
			0 & 1 & 2 & 2 & 0.046 & 0.045 & 0.019 & \cellcolor{yellow} 0.017 & 0.033 & 0.048 \\ \hline
			1 & 9 & 15 & 3 & \cellcolor{gray} ER1 & \cellcolor{gray} TO & \cellcolor{gray} TO & \cellcolor{gray} TO & \cellcolor{gray} TO & \cellcolor{gray} TO \\ \hline
			2 & 6 & 8 & 3 & 0.704 & \cellcolor{yellow} 0.11 & \cellcolor{gray} GU & \cellcolor{gray} GU & 0.541 & 14.65 \\ \hline
			3 & 4 & 8 & 3 & \cellcolor{yellow} 0.013 & 0.034 & \cellcolor{gray} GU & \cellcolor{gray} GU & 0.101 & 0.12 \\ \hline
			4 & 2 & 6 & 2 & 0.007 & 0.032 & \cellcolor{yellow} 0.0 & 0.016 & 0.017 & 0.036 \\ \hline
			5 & 2 & 6 & 3 & \cellcolor{yellow} 0.012 & 0.062 & 0.063 & 0.063 & 0.03 & 0.049 \\ \hline
			6 & 2 & 6 & 3 & \cellcolor{yellow} 0.01 & 0.016 & \cellcolor{gray} GU & \cellcolor{gray} GU & 0.049 & 0.035 \\ \hline
			7 & 2 & 6 & 5 & \cellcolor{yellow} 0.007 & 0.032 & 0.016 & 0.008 & 0.026 & 0.044 \\ \hline
			8 & 3 & 8 & 3 & \cellcolor{yellow} 0.013 & 0.022 & \cellcolor{gray} GU & \cellcolor{gray} GU & 0.085 & 0.07 \\ \hline
			9 & 3 & 8 & 3 & \cellcolor{yellow} 0.016 & 0.032 & \cellcolor{gray} GU & \cellcolor{gray} GU & 0.078 & 0.074 \\ \hline
			10 & 3 & 8 & 3 & \cellcolor{yellow} 0.016 & 0.032 & \cellcolor{gray} GU & \cellcolor{gray} GU & 0.095 & 0.083 \\ \hline
			11 & 3 & 9 & 3 & 14.69 & \cellcolor{yellow} 0.05 & \cellcolor{gray} GU & \cellcolor{gray} GU & 0.16 & 0.104 \\ \hline
			12 & 3 & 9 & 3 & \cellcolor{yellow} 0.004 & 0.047 & 0.081 & 0.094 & 0.096 & 0.065 \\ \hline
			13 & 3 & 9 & 3 & \cellcolor{yellow} 0.015 & 0.03 & \cellcolor{gray} GU & \cellcolor{gray} GU & 0.114 & 0.082 \\ \hline
			14 & 3 & 10 & 3 & \cellcolor{gray} TO & \cellcolor{yellow} 0.751 & 500.7 & 430.5 & 11.19 & \cellcolor{gray} TO \\ \hline
			15 & 3 & 10 & 4 & \cellcolor{yellow} 0.04 & 0.141 & \cellcolor{gray} GU & \cellcolor{gray} GU & 0.112 & 0.116 \\ \hline
			16 & 4 & 11 & 4 & \cellcolor{gray} TO & 61.13 & \cellcolor{gray} GU & \cellcolor{gray} GU & \cellcolor{yellow} 1.272 & 9.846 \\ \hline
			17 & 4 & 11 & 3 & \cellcolor{gray} TO & 30.74 & \cellcolor{gray} TO & 298.8 & \cellcolor{gray} TO & \cellcolor{yellow} 5.056 \\ \hline
			18 & 4 & 12 & 2 & \cellcolor{yellow} 0.011 & 0.046 & \cellcolor{gray} GU & \cellcolor{gray} GU & 0.133 & 0.122 \\ \hline
			19 & 4 & 12 & 3 & 0.017 & \cellcolor{yellow} 0.016 & \cellcolor{gray} GU & \cellcolor{gray} GU & 0.11 & 0.15 \\ \hline
			20 & 4 & 12 & 3 & 0.174 & \cellcolor{yellow} 0.108 & \cellcolor{gray} GU & \cellcolor{gray} GU & 0.332 & 0.203 \\ \hline
			21 & 4 & 13 & 5 & \cellcolor{yellow} 0.016 & 0.031 & \cellcolor{gray} GU & \cellcolor{gray} GU & 0.216 & 0.169 \\ \hline
			22 & 5 & 13 & 3 & 0.191 & \cellcolor{yellow} 0.047 & \cellcolor{gray} GU & \cellcolor{gray} GU & 0.676 & 0.37 \\ \hline
			23 & 5 & 13 & 4 & \cellcolor{gray} TO & \cellcolor{gray} TO & \cellcolor{gray} GU & \cellcolor{gray} GU & \cellcolor{gray} TO & \cellcolor{gray} TO \\ \hline
			24 & 6 & 15 & 3 & \cellcolor{gray} TO & 13.12 & \cellcolor{gray} GU & \cellcolor{gray} GU & \cellcolor{yellow} 1.649 & 2.507 \\ \hline
			25 & 3 & 9 & 3 & \cellcolor{gray} TO & 0.256 & 1.849 & 1.741 & 0.247 & \cellcolor{yellow} 0.192 \\ \hline
			26 & 4 & 10 & 3 & \cellcolor{gray} TO & \cellcolor{yellow} 0.207 & \cellcolor{gray} GU & \cellcolor{gray} GU & 0.407 & 0.337 \\ \hline
			27 & 4 & 11 & 3 & \cellcolor{gray} TO & \cellcolor{yellow} 39.89 & \cellcolor{gray} GU & \cellcolor{gray} GU & \cellcolor{gray} TO & 201.6 \\ \hline
			28 & 4 & 11 & 4 & \cellcolor{gray} TO & 30.72 & \cellcolor{gray} GU & \cellcolor{gray} GU & \cellcolor{gray} TO & \cellcolor{yellow} 14.62 \\ \hline
			29 & 4 & 12 & 3 & \cellcolor{gray} TO & 4.169 & \cellcolor{gray} GU & \cellcolor{gray} GU & \cellcolor{yellow} 1.435 & 8.0 \\ \hline
			30 & 5 & 13 & 3 & \cellcolor{gray} TO & \cellcolor{yellow} 8.447 & \cellcolor{gray} GU & \cellcolor{gray} GU & \cellcolor{gray} TO & 279.9 \\ \hline
			31 & 5 & 13 & 4 & \cellcolor{gray} TO & \cellcolor{gray} TO & \cellcolor{gray} GU & \cellcolor{gray} GU & \cellcolor{gray} TO & \cellcolor{gray} TO \\ \hline
			32 & 5 & 14 & 4 & \cellcolor{gray} ER1 & \cellcolor{gray} TO & \cellcolor{gray} GU & \cellcolor{gray} GU & \cellcolor{gray} TO & \cellcolor{gray} TO \\ \hline
			33 & 5 & 14 & 3 & \cellcolor{gray} TO & \cellcolor{gray} TO & \cellcolor{gray} GU & \cellcolor{gray} GU & \cellcolor{gray} TO & \cellcolor{gray} TO \\ \hline
			34 & 5 & 14 & 3 & \cellcolor{gray} ER1 & 546.9 & \cellcolor{gray} GU & \cellcolor{gray} GU & \cellcolor{yellow} 27.93 & \cellcolor{gray} TO \\ \hline
			35 & 6 & 18 & 4 & \cellcolor{gray} ER1 & \cellcolor{gray} TO & \cellcolor{gray} GU & \cellcolor{gray} GU & \cellcolor{gray} TO & \cellcolor{gray} TO \\ \hline
			36 & 2 & 2 & 3 & 0.07 & 0.154 & \cellcolor{yellow} 0.04 & 0.047 & 0.061 & 0.074 \\ \hline
			37 & 3 & 2 & 3 & \cellcolor{gray} TO & 453.7 & 0.445 & 0.493 & 0.618 & \cellcolor{yellow} 0.361 \\ \hline
			38 & 4 & 2 & 3 & \cellcolor{gray} TO & \cellcolor{gray} TO & \cellcolor{gray} GU & \cellcolor{gray} GU & 1.581 & \cellcolor{yellow} 0.977 \\ \hline
			39 & 5 & 2 & 3 & \cellcolor{gray} TO & \cellcolor{gray} TO & \cellcolor{gray} GU & \cellcolor{gray} GU & 14.03 & \cellcolor{yellow} 11.53 \\ \hline
			40 & 6 & 2 & 3 & \cellcolor{gray} TO & \cellcolor{gray} TO & \cellcolor{gray} GU & \cellcolor{gray} GU & \cellcolor{yellow} 155.6 & 529.0 \\ \hline
			41 & 7 & 2 & 3 & \cellcolor{gray} TO & \cellcolor{gray} TO & \cellcolor{gray} GU & \cellcolor{gray} GU & \cellcolor{gray} ER2 & \cellcolor{gray} TO \\ \hline
			42 & 2 & 4 & 4 & \cellcolor{gray} TO & \cellcolor{gray} TO & 243.6 & 240.4 & \cellcolor{yellow} 100.9 & 240.7 \\ \hline
			43 & 2 & 1 & 2 & 0.027 & 0.016 & 0.019 & \cellcolor{yellow} 0.0 & 0.042 & 0.042 \\ \hline
			44 & 3 & 1 & 2 & \cellcolor{yellow} 0.01 & 0.095 & 0.033 & 0.04 & 0.088 & 0.088 \\ \hline
			45 & 4 & 1 & 2 & \cellcolor{yellow} 0.047 & 1.991 & \cellcolor{gray} GU & \cellcolor{gray} GU & 0.187 & 0.16 \\ \hline
			46 & 5 & 1 & 2 & \cellcolor{yellow} 0.078 & 72.69 & \cellcolor{gray} GU & \cellcolor{gray} GU & \cellcolor{gray} ER2 & 0.4 \\ \hline
			47 & 6 & 1 & 2 & \cellcolor{yellow} 0.294 & \cellcolor{gray} TO & \cellcolor{gray} GU & \cellcolor{gray} GU & \cellcolor{gray} ER2 & 1.099 \\ \hline
			48 & 7 & 1 & 2 & \cellcolor{yellow} 2.361 & \cellcolor{gray} TO & \cellcolor{gray} GU & \cellcolor{gray} GU & \cellcolor{gray} ER2 & 3.026 \\ \hline
		\end{tabular}
	}
	\caption{Benchmarking of the six approaches for 49 systems coming from CRN theory and population dynamics. For a full report see the report spreadsheet file in the repository of the paper.\label{tab:benchmark}}
\end{table}

\section{Conclusion}
\label{sec:Conclusion}

In this paper we introduced new methods to decompose the parameter space into regions where a parametric system of polynomial equations has different numbers of solutions.  The prior state-of-the-art has had worst case doubly exponential complexity in both its first and second parts, whereas the new algorithms reduce the doubly exponential growth in the number of polynomials in the first part; to somewhere between singly and doubly exponential. The benefits of the new approaches were validated through the symbolic solution of a real world example in a few minutes which was infeasible for the prior state-of-the-art.  This also showed that the observations in \cite{Rost-Sadeghimanesh-2021-1} found by a symbolic-numeric algorithm were indeed a complete classification of the dynamical behaviour of the system.  

Our benchmarking shows the applicability of our new approaches in the general case of problems from population dynamics and CRN theory that fit the problem description in Section~\ref{sec:Problem_Statement}. It also provided examples for different scenarios where one approach is better than the rest and therefore rejecting the hypothesis that any of the six approaches can be always better than the rest.  This suggests an avenue of future work in algorithm selection for these problems, and we note the recent work using machine learning for this \cite{BEG24b,BEG25,BSEG24}.

To tackle larger examples, one option would be to combine the new symbolic approaches with numerical sampling, as was done in \cite{Rost-Sadeghimanesh-2021-1}.  We intend to explore the limit of the new algorithms in hybrid as future work.  There is also hope to improve the symbolic methods as future work:  both by the use of additional projection technology to deal with the cases of \texttt{ResChainBranching} where a branch has a single polynomial; and by further optimisations to remove the remaining redundancies in the output of \Cref{Alg:ResChainBranching}.

\subsubsection*{Data Access Statement}

The code and data described in this paper is openly available from this URL: \url{https://doi.org/10.5281/zenodo.22860460}

\subsubsection*{Acknowledgements}

The authors acknowledge the support of EPSRC Grant EP/T015748/1, ``\textit{Pushing Back the Doubly-Exponential Wall of Cylindrical Algebraic Decomposition}''. We thank Tereso del R\'io for useful conversations, and the anonymous reviewers of an earlier version of this paper for helpful feedback.

\newpage

\FloatBarrier


\end{document}